# Towards Control of Dam and Reservoir Systems with Forward-Backward Stochastic Differential Equations Driven by Clustered Jumps

**(Short title: Control of Dam and Reservoir Systems with FBSDEs)**

Hidekazu Yoshioka[1, 2, *]

[1]Graduate School of Natural Science and Technology, Shimane University, 1060 Nishikawatsu, Matsue, 690-8504, Japan

[2]Fisheries Ecosystem Project Center, Shimane University, 1060 Nishikawatsu, Matsue, 690-8504, Japan

*Corresponding author: yoshih@life.shimane-u.ac.jp

**Abstract**

We deal with a new maximum principle-based stochastic control model for river management through operating a dam and reservoir system. The model is based on coupled forward-backward stochastic differential equations (FBSDEs) derived from jump-driven streamflow dynamics and reservoir water balance. A continuous-time branching process with immigration driven by a tempered stable subordinator efficiently describes clustered inflow streamflow dynamics. This is a completely new attempt in hydrology and control engineering. Applying a stochastic maximum principle to the dynamics based on an objective functional for designing cost-efficient control of dam and reservoir systems leads to the FBSDEs as a system of optimality equations. The FBSDEs under a linear-quadratic ansatz lead to a tractable model, while they are solved numerically in the other cases using a least-squares Monte-Carlo method. Optimal controls are found in the former, while only sub-optimal ones are computable in the latter due to a hard state constraint. Model parameters are successfully identified from a real data of a river in Japan having a dam and reservoir system. We also show that the linear-quadratic case can capture the real operation data of the system with underestimation of the outflow discharge. More complex cases with a realistic time horizon are analyzed numerically to investigate impacts of considering the environmental flows and seasonal operational purposes. Key challenges towards more sophisticated modeling and analysis with jump-driven FBSDEs are discussed as well.

## 1. Introduction

### 1.1 Study background

Living together with aquatic environment and ecosystems is a crucial issue. In particular, human interventions to river environment, including construction of dams and weirs [1-2], water abstraction [3-4], and environmental pollution [5-6] triggered severe problems that consequently affect human lives in a variety of forms such as the decline of fish catches [7-8]. In this paper, we focus on dam and reservoir systems as elements shaping river environment under anthropogenic pressure [9]. Constructing a dam in a river not only leads to hydraulic alternation of its downstream rivers but also affects its upstream reach because the flowing water is stored in the reservoir thereby creating an artificial lentic environment [10].

For the downstream rivers, outflow discharges from a dam and/or its fluctuation should not be too small to suppress bloom of nuisance benthic algae and vegetation [11-12]. This is a requirement of discharge called environmental flows and has been a key concept in operating dam and reservoir systems in the modern world [13]. By contrast, too large outflow should be avoided for flood preventions [14]. Water temperature as a metric of habitat quality in a dam-downstream river is critically affected by dam operations [15]. While for the upstream reservoirs, water levels (or water depths) and their speed of change affect aquatic species living in them. In the Three Gorges Reservoir, spawning areas of carp species depend on the water level where low water level is preferred for human lives while high water levels are required to sustain fish population [16]. Bivalve populations are constrained by annual water level drawdowns [17]. Rapid water level changes of a reservoir perish fish species living in it while some other species increased abundance, suggesting that changing water levels serves as a new technique to control fish assemblages in reservoirs [18-20]. Operation purpose of dam and reservoir systems sometimes include hydropower generation and water quality management as well [21-22].

### 1.2 Mathematical background

As reviewed above, operation of a dam and reservoir system should balance human interventions against aquatic environment and ecosystems as well as their conservation and restoration. This engineering problem can be understood as a control problem. It is commonly assumed that streamflow dynamics is stochastic phenomena driven by climatic, environmental, and anthropogenic factors that are highly complicated and often unpredictable [23-24].

Controlling stochastic dynamics can be approached from the standpoint of control of stochastic differential equations (SDEs) based on maximum or dynamic programming principle [25]. The main advantage of the stochastic control approaches is the flexibility to deal with a wide class of dynamics driven by continuous and jump noises. Applications of stochastic control are found in a variety of research areas including finance [26], insurance [27], management of resources and environment [28-29], biological and decease control [30-31], and modern industries like software testing [32] and cyber-physical design [33].

Maximum principle and dynamic programming principle are formally equivalent [25]. The dynamic programming principle reduces solving a stochastic control problem to finding a solution to a degenerate parabolic partial differential equation called Hamilton-Jacobi-Bellman equation. These equations can only reasonably manage problems having one or two state variables [34-36] unless a special approximation technique to deal with the curse of dimensionality is used [37-38]. The stochastic maximum principle, by contrast, reduces resolving a control problem to solving a system of forward-backward stochastic differential equations (FBSDEs) applicable to higher dimensions [39-41]. Another difference between the two approaches is that simulating both control and controlled dynamics is naturally carried out in the latter, while they must be addressed separately in the former. We focus on the maximum principle because our interests are in both optimal control and controlled dynamics.

Recent studies suggest that streamflow time series as a driving noise process of water balance in a dam and reservoir system follows jump-driven SDEs [23]. Markov chain models as discretized SDEs have also been proposed [42]. By using an SDE model of streamflow time series, designing operation rules of a dam and reservoir system reduces to solving a stochastic control problem. The maximum principle would then be preferred because there may exist several state variables to be considered, including inflow and outflow discharges, reservoir water level, and further water quality and biological indices in some cases [23, 43-44]. Such an approach is still rare to the best of the author's knowledge despite their importance.

### 1.3 Objectives and contributions

The objectives of this paper are formulation, analysis, and application of a new stochastic control model of dam and reservoir system based on FBSDEs. To present key challenges in numerical computation of jump-driven FBSDEs is also an objective of this paper. Our model is simple, but it advances modeling and control of dam and reservoir systems as explained below.

Our target system has the three state variables: inflow discharge to a dam and reservoir system, reservoir water volume, and outflow discharge from the system. We assume that the outflow is controlled by modulating its speed of change, motivated by real operation data of a dam and reservoir system that the outflow discharge is regulated by controlling the system. Controlling the change of speed of outflow in the proposed way is reasonable from an engineering viewpoint. This is because sudden changes of river discharge and thus water level should be avoided since it threatens both upstream reservoir and downstream rivers [45-46]. Our model is parsimonious compared with those in the application studies since it has a fewer degree-of-freedom [47-49] but opens a new control viewpoint of modeling dam and reservoir systems.

A characteristic of streamflow is that the discharge at a fixed point in a river has clustered spikes in time because of the self-exciting nature of the weather generating mechanism seen in real cases [50-51]. Such a clustering mechanism, ubiquitous in hydrology [52], cannot be considered using the conventional Lévy processes because of having independent increments. To overcome this difficulty efficiently, we use a continuous branching process with immigration, CBI process in short [53]. Using a real data set, we show that the CBI process driven by tempered stable jumps, previously used in financial modeling [54] and power market dynamics [55], is a simple candidate SDE model generating streamflow time series. This is a new use of CBI processes in engineering research areas.

Our objective functional as an index to be maximized is based on a linear-quadratic (LQ) one aiming at achieving targeted outflow discharge and water volume, but has a non-quadratic term considering environmental flows. Although the final target in this paper is non-LQ problems focusing on environmental flows, the LQ case is useful because it is solvable exactly as well as giving insights into better understanding the non-LQ case as in the literature [56-57]. Stochastic control, especially maximum principle with CBI processes, is still a germinating topic [31] and our contribution provides a primitive example.

Based on a real operation data, we show that the LQ case reasonably capture characteristics of the real data. In the non-LQ case, state constraints of the dynamics, such as non-negativity of the outflow discharge and water volume, are additionally considered. Control with state constraints is well-studied for dynamic programming [42, 58-59]. However, it has not been well-studied for maximum principle; especially, the boundary discontinuity of the admissible range of controls hinders us from directly applying it to solving optimal control problems. We tackle this issue by considering sub-optimal controls imposing the constraint only in the forward dynamics [60]. Sub-optimal controls can be used effectively if they are feasible and more easily computable than the optimal ones [61-62]; both are satisfied in our case.

We solve the FBSDEs numerically. Our numerical method uses a least-squares Monte-Carlo method [40] for FBSDEs enhanced with the stochastic grid bundling [63] to adaptively control computational resolution under a given computational resource. The stochastic grid bundling, to the best of the author's knowledge, has been applied to decoupled FBSDEs but not to coupled FBSDEs. The relaxed Picard iteration [64] enables us to iteratively compute the discretized FBSDEs in a stable manner. The sub-optimal approach is justified in our application because of the small probability of the controlled states staying close to the state boundaries.

Consequently, we contribute to new mathematical modeling of dam and reservoir systems based on CBI processes and FBSDEs and their application. The problem we deal with is simple but potentially serves as a building block for advanced models that are more complicated and realistic in future.

### 1.4 Structure of this paper

In **Section 2**, the target dynamics and the associated FBSDEs are presented. In **Section 3**, we give a sufficient maximum principle of the FBSDEs and show that a LQ case is solvable analytically. More complicated cases are considered in **Section 4** where the FBSDEs are computed numerically using an identified model. In **Section 5**, summary of this paper is presented and future directions of our research is discussed. **Appendix A** explains a preliminary finding that motivated the proposed model. **Appendix B** explains an analytical technique to calculate moments of CBI processes. Proofs of the propositions in the maim text are placed in **Appendix C**.

## 2. System dynamics and optimality

### 2.1 Target dynamics

We work with a continuous-time setting under a usual complete probability space [25]. Consider a dam and reservoir system receiving an inflow process $Q=(Q_t)_{t\geq 0}$ that is càdlàg (right continuous with left limits) while releasing controllable outflow $q=(q_t)_{t\geq 0}$ (**Figure 1**). Another state variable is the water volume $V=(V_t)_{t\geq 0}$ evolving according to the inflow and outflow. The speed of change of the outflow is denoted as $a=(a_t)_{t\geq 0}$, which is the control process in the proposed model. The reservoir capacity is $\bar{V}>0$.

The governing SDEs of the three state variables are as follows: the SDE of inflow

$$\begin{aligned} \mathrm{d}Q_t &= \rho\left(\underline{Q}-Q_t\right)\mathrm{d}t+\int_0^{\infty}\int_0^{Q_{t-}} zN\left(\mathrm{d}u,\mathrm{d}z,\mathrm{d}t\right) \\ &\left(=\rho\left\{\underline{Q}-(1-M_1)Q\right\}\mathrm{d}t+\int_0^{\infty}\int_0^{Q_{t-}} z\tilde{N}\left(\mathrm{d}u,\mathrm{d}z,\mathrm{d}t\right)\right) \end{aligned} \quad \text{for } t>0,\ Q_0\geq 0, \tag{1}$$

the SDE of outflow

$$\mathrm{d}q_t = a_t\mathrm{d}t \quad \text{for } t>0,\ q_0\geq 0, \tag{2}$$

and the SDE of reservoir water balance

$$\mathrm{d}V_t = \begin{cases} \left(Q_t-q_t\right)\mathrm{d}t & \text{(Unconstrained case)} \\ \left(Q_t-q_t\right)\mathrm{d}t+\mathrm{d}\underline{\eta}_t-\mathrm{d}\bar{\eta}_t & \text{(Constrained case)} \end{cases} \quad \text{for } t>0,\ V_0\geq 0. \tag{3}$$

Each component of the system is explained in what follows. The SDE (1) is a CBI process driven by a state-dependent jump noise and a linear drift term. Here, $\underline{Q}>0$ is the minimum inflow discharge, $\rho>0$ is the recession rate, and $N\left(\mathrm{d}u,\mathrm{d}z,\mathrm{d}t\right)$ is a space-time Poisson random measure on $(0,+\infty)^3$ with the compensator $\mathrm{d}uv\left(\mathrm{d}z\right)\mathrm{d}t$ and the compensated measure [53] $\tilde{N}\left(\mathrm{d}u,\mathrm{d}z,\mathrm{d}t\right)=N\left(\mathrm{d}u,\mathrm{d}z,\mathrm{d}t\right)-\mathrm{d}uv\left(\mathrm{d}z\right)\mathrm{d}t$, where $v\left(\mathrm{d}z\right)$ is the integral kernel of the finite-variation tempered stable type [54]:

$$v\left(\mathrm{d}z\right)=\frac{\rho a}{z^{1+\alpha}}e^{-bz}\mathrm{d}z \tag{4}$$

with the parameters $a>0$, $b>0$, and $\alpha\in(0,1)$ representing frequency, strength, and intermittency of jumps exciting the inflow $Q$, respectively. We assume $M_1=ab^{\alpha-1}\Gamma(1-\alpha)\in(0,1)$ so that

$$\rho\left(1-M_1\right)=\rho\left(1-ab^{\alpha-1}\Gamma(1-\alpha)\right)>0. \tag{5}$$

The expectation $\mathbb{E}\left[Q_t\right]$ of $Q_t$, the average inflow, satisfies the ordinary differential equation (ODE):

$$\frac{\mathrm{d}}{\mathrm{d}t}\mathbb{E}\left[Q_t\right]=\rho\left\{\underline{Q}-\left(1-M_1\right)\mathbb{E}\left[Q_t\right]\right\} \quad \text{for } t>0, \tag{6}$$

meaning that $\lim_{t\to+\infty}\mathbb{E}\left[Q_t\right]=\left(1-M_1\right)^{-1}\underline{Q}>0$ if (5) is satisfied while $\lim_{t\to+\infty}\mathbb{E}\left[Q_t\right]=+\infty$ otherwise. Physically, flow discharge in a river should be bounded in the mean, which is the reason we assume (5).

Because of the CBI nature, the inflow process can be seen as an autoregressive process driven by a jump process having the state-dependent kernel $\frac{Q_{t-}\rho a}{z^{1+\alpha}}e^{-bz}\mathrm{d}z$. An intuitive explanation is that a larger inflow discharge corresponding to a flood event leads to the larger (formal) Lévy measure and thus more frequent jumps. This self-exciting mechanism generates self-exciting jumps corresponding to clustered flood events like those found in rainy season and snow-melting seasons. A natural filtration generated by the random measure $N$ up to time $t$ is denoted as $\mathcal{F}_t$. Set $\mathcal{F}=(\mathcal{F}_t)_{t\geq 0}$.

The SDE (2) represents the physical law that the outflow is obtained by integrating its speed of change. Finally, the SDE (3) represents the physical law that the water balance is determined by the difference between inflow and outflow, where the other factors such as evaporation from the water surface and filtration into surrounding groundwater are not considered to simplify the model, but can be considered if necessary as additional source/sink terms.

The constrained and unconstrained cases of the SDE (3) are explained separately in the next sub-section. We firstly discuss the unconstrained case, and then explain the constrained case by emphasizing differences between the two cases. The unconstrained case is simpler case where the state variables $(q,V)$ are allowed to be valued in $\mathbb{R}^2$, while they are constrained in a compact set in the constrained case.

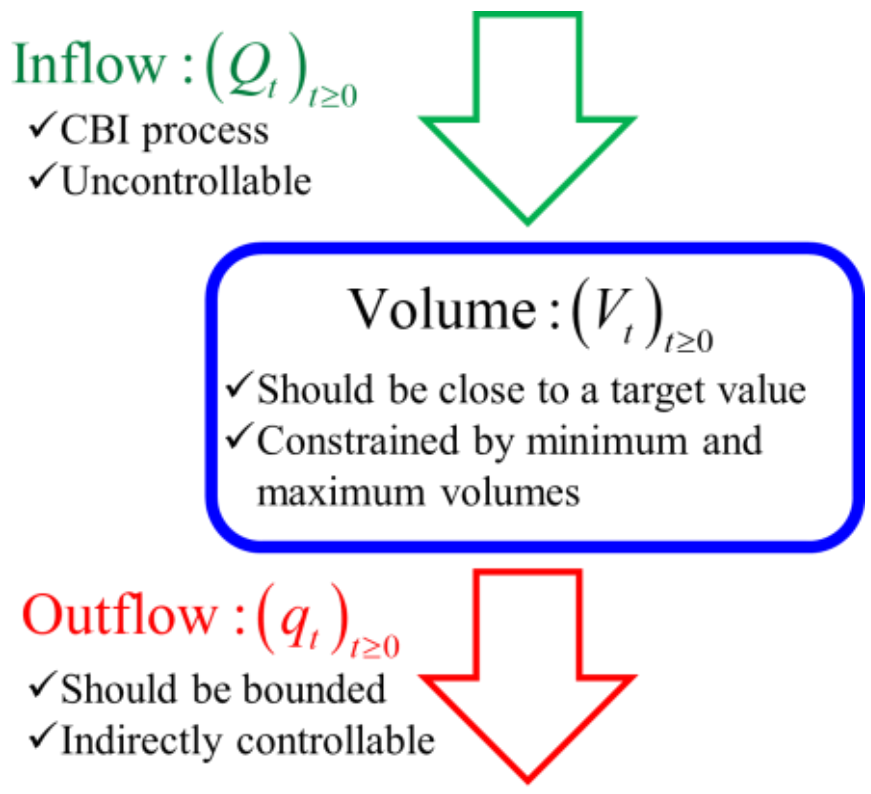

**Figure 1.** A conceptual figure of the dam and reservoir system.

## 2.2 State constraint

Our system has the three state variables where only $q$ and $V$ are controllable. Indeed, the SDE (1) does not contain $q$, $V$, and $a$. By Theorem 3.1 of Dawson and Li [53], the SDE (1) has a unique path-wise càdlàg solution globally in time and the solution is non-negative. It is therefore well-defined. By contrast, treatment of the state variables $q$ and $V$ are more complicated because they should be constrained in compact sets $\Omega_V = [0,\bar{V}]$ and $\Omega_q = [0,\bar{q}]$ with $\bar{q} > 0$, respectively. The constraint $V_t \in \Omega_V$ ($t \geq 0$) arises because the capacity of a reservoir should be finite and the stored water volume should be non-negative as well as not larger than the capacity. The constraint $q_t \in \Omega_q$ ($t \geq 0$) arises because the outflow discharge should be non-negative and there will be a technological limit constraining its maximum. Each admissible $a = (a_t)_{t\geq 0}$ should be chosen so that the state variables are constrained in proper sets.

For the constrained case, the process $\underline{\eta} = (\underline{\eta}_t)_{t\geq 0}$ ($t \geq 0$) is non-increasing and càdlàg with $\int_0^t V_s \mathrm{d}\underline{\eta}_s = 0$. Similarly, the process $\bar{\eta} = (\bar{\eta}_t)_{t\geq 0}$ ($t \geq 0$) is non-increasing and càdlàg with $\int_0^t (V_s - \bar{V}) \mathrm{d}\underline{\eta}_s = 0$. These processes, which are absent in the unconstrained case, represent emergency operations of the system to constrain the water volume in $\Omega_V = [0, \bar{V}]$. Employing the singular processes means projecting the water volume $V_t$ into the $\Omega_V$ at each $t$. The constraints are considered as boundary reflections of state variables as discussed in Xu et al. [66] and Ghosh and Pradhan [67]. For readers' sake, we again emphasize that the range of the couple $(q_t, V_t)$ is $\mathbb{R}^2$ for the unconstrained case and $\Omega_q \times \Omega_V$ for the constrained case.

***Remark 1*** There may exist other ways to constrain the water volume by incorporating càdlàg processes into the dynamics of $q$ and activating them when $V_t = 0$ and $V_t = \bar{V}$. One may also consider a stopping time to terminate the problem where the problem is terminated if $V_t = 0$ or $V_t = \bar{V}$. Our state-constraint problems consider neither such a termination nor penalization. These problems will concern safety aspects of dam operations because the boundary states, full and empty of the reservoir volume, should be avoided. They are not considered here because our problem is already non-trivial and will be addressed future.

We introduce two admissible sets of the control $a$. The first set $\mathcal{A}_1$ is

$$\mathcal{A}_1 = \left\{ (a_t)_{t\geq 0} \,\middle|\, \begin{array}{l} a_t \text{ is progressibly measurable with respect to } \mathcal{F}_t \text{ and } \int_0^t a_t^2 \mathrm{d}s < +\infty \\ \text{and bounded in } A = [-\bar{a}, \bar{a}] \text{ for each } t \geq 0, \text{ and } (q_t, V_t)_{t\geq 0} \text{ are determined uniquely.} \end{array} \right\}, \quad (7)$$

where $\bar{a} > 0$ and we set $A = \mathbb{R}$ when $\bar{a} = +\infty$. In this case, the triplet $(Q_t, q_t, V_t)$ has the range $\mathbb{R}^3$. The square integrability $\int_0^t a_t^2 \mathrm{d}s < +\infty$ automatically follows if $\bar{a} < +\infty$. Disregarding the state variables implies the occurrence of unphysical states like $V_t < 0$ and $q_t < 0$. This case can be considered as an approximation of the constrained case explained below if the state variables satisfy the constraint with high probability.

The second admissible set $\mathcal{A}_2$ is

$$\mathcal{A}_2 = \left\{ (a_t)_{t\geq 0} \,\middle|\, a \in \mathcal{A}_1 \text{ and } (q_t, V_t) \in \Omega_q \times \Omega_V \ (t \geq 0). \right\}. \quad (8)$$

This is a physically more reasonable version of $\mathcal{A}_1$. We introduce the state-dependent range $A(q_t)$ of the controls complying with (8):

$$A(q_t) = \begin{cases} [-\bar{a}, 0] & (q_t = \bar{q}) \\ [0, \bar{a}] & (q_t = 0) \\ [-\bar{a}, \bar{a}] & (\text{Otherwise}) \end{cases}. \quad (9)$$

Hereafter, $\mathcal{A}$ represents either $\mathcal{A}_1$ or $\mathcal{A}_2$ and will be specified when necessary. The former is used in the unconstrained case, while the latter in the constrained case.

### 2.3 Objective functional

The objective functional is an index to be maximized with respect to $a \in \mathcal{A}$. Set the conditional expectation $\phi$ as an objective functional:

$$\phi(t,Q_t,q_t,V_t;a)=\mathbb{E}\left[\int_t^T e^{-\delta(s-t)}J(s,Q_s,q_s,V_s,a_s)\mathrm{d}s\middle|\mathcal{F}_t\right] \tag{10}$$

with

$$-J(s,Q_s,q_s,V_s,a_s)=\frac{w_1}{2}(Q_s-q_s)^2+\frac{w_2}{2}\left(V_s-\hat{V}_s\right)^2+\frac{w_3}{2}a_s^2+j(q_s). \tag{11}$$

Here, $T>0$ is a prescribed terminal time and $\delta>0$ is a discount rate representing how myopic the decision-maker (the system operator) is: larger $\delta$ represents the decision-maker whose weights more on current state than future states. The constants $w_1,w_2,w_3>0$ are weighting constants and $j:\mathbb{R}\to\mathbb{R}$ is a smooth convex function that growths at most quadratically. This growth bound is necessary for the sufficient maximum principle (**Proposition 1**).

The (negative of) first through fourth terms in the right-hand side of (11) correspond to a penalization from the run-of-river condition where the inflow should be close to outflow [24, 42, 68], a penalization of the water volume from a prescribed piecewise smooth target $\left(\hat{V}_t\right)_{t\geq 0}$, and a penalization of fast variation of $q$, and an additional penalization later utilized for constraining too large and/or too small outflow discharges that may critically disturb the downstream river environment. For a motivation of the proposed objective functional, please see preliminary finding of **Appendix A**. In **Section 4**, we use $j(q)=\frac{w_4}{2}\max\left\{\underline{q}-q,0\right\}^2$ with $w_4>0$ and $\underline{q}>0$ softly constraining the outflow. This is the simplest penalty model of environmental flows [24].

The value function as the optimized $\phi$ with respect to $a\in\mathcal{A}$ is defined as

$$\Phi(t,Q_t,q_t,V_t)=\sup_{a\in\mathcal{A}}\phi(t,Q_t,q_t,V_t;a), \tag{12}$$

where the left-hand side is a space-time 4-D function having the arguments $(t,Q_t,q_t,V_t)$. The goal of our control problem is to find a maximizer $a^*\in\mathcal{A}$ of (12). We are interested in feedback Markovian optimal controls of the form $a_t^*=a^*(t,Q_t,q_t,V_t)$ (with an abuse of notations as in conventional problems [25].)

## 2.4 FBSDEs

### 2.4.1 Unconstrained case

By using FBSDEs, we solve the control problem without directly resorting to the maximization (12). We reduce the control problem to FBSDEs using a stochastic maximum principle. The classical maximum principle needs to be modified because it does not cover SDEs driven by CBI processes. Hess [31] presented a maximum principle with CBI processes, which also applies to our unconstrained case $\mathcal{A}=\mathcal{A}_1$.

Set the Hamiltonian

$$H=H\left(t,Q,q,V,p^{(Q)},p^{(q)},p^{(V)},\theta^{(Q)},\theta^{(Q)},\theta^{(Q)},a\right):[0,T]\times\mathbb{R}^3\times\mathbb{R}^3\times\mathfrak{A}\left(\mathbb{R}_+^2\right)^3\times[-\bar{a},\bar{a}]\to\mathbb{R} \tag{13}$$

by

$$
\begin{aligned}
H =& -\left\{\frac{w_1}{2}(q-Q)^2+\frac{w_2}{2}\left(V-\hat{V}_t\right)^2+\frac{w_3}{2}a^2+j(q)\right\} \\
&+\rho\left\{\underline{Q}-(1-M_1)Q\right\}p^{(Q)}+ap^{(q)}+(Q-q)p^{(V)} \\
&+\int_0^\infty\int_0^Q\left[\theta^{(Q)}(u,z)+\theta^{(q)}(u,z)+\theta^{(V)}(u,z)\right]\mathrm{d}u\nu(\mathrm{d}z) \\
&-\delta\left(Qp^{(Q)}+qp^{(q)}+Vp^{(V)}\right)
\end{aligned}
\tag{14}
$$

with $\mathfrak{A}$ a space of functions with which the integral of (14) converges [31]. We have

$$
\frac{\partial H}{\partial a}=-w_3a+p^{(q)} \quad \text{and} \quad \frac{\partial^2 H}{\partial a^2}=-w_3<0\,,
\tag{15}
$$

leading to that $H$ as a function of $a$ is maximized at

$$
\hat{a}\left(p^{(q)}\right)=\max\left\{-\bar{a},\min\left\{\bar{a},\frac{p^{(q)}}{w_3}\right\}\right\}.
\tag{16}
$$

The other partial derivatives of $H$ used in what follows are

$$
\frac{\partial H}{\partial Q}=-w_1(Q-q)-\left(\rho(1-M_1)+\delta\right)p^{(Q)}+p^{(V)}+\int_0^\infty\left[\theta^{(Q)}(Q,z)+\theta^{(q)}(Q,z)+\theta^{(V)}(Q,z)\right]\nu(\mathrm{d}z)\,,
\tag{17}
$$

$$
\frac{\partial H}{\partial q}=-w_1(q-Q)-\frac{\mathrm{d}j(q)}{\mathrm{d}q}-\delta p^{(q)}-p^{(V)}\,,
\tag{18}
$$

$$
\frac{\partial H}{\partial V}=-w_2\left(V-\hat{V}_t\right)-\delta p^{(V)}\,.
\tag{19}
$$

Consider the càdlàg triplet $\left(p_t^{(Q)},p_t^{(q)},p_t^{(V)}\right)_{t\geq 0}$ and the triplet $\left(\theta^{(Q)}(t,\cdot,\cdot),\theta^{(q)}(t,\cdot,\cdot),\theta^{(V)}(t,\cdot,\cdot)\right)_{t\geq 0}$ of càdlàg time-dependent maps from $\mathbb{R}_+^2$ to $\mathbb{R}$. Set the adjoint system (or backward stochastic differential equations, BSDEs in short) governing $\left(p_t^{(Q)},p_t^{(q)},p_t^{(V)},\theta^{(Q)}(t,\cdot,\cdot),\theta^{(q)}(t,\cdot,\cdot),\theta^{(V)}(t,\cdot,\cdot)\right)_{t\geq 0}$:

$$
\begin{aligned}
\mathrm{d}p_t^{(Q)} &= -\frac{\partial H}{\partial Q}\mathrm{d}t+\int_0^\infty\int_0^{Q_{t-}}\theta^{(Q)}(t-,u,z)\tilde{N}(\mathrm{d}u,\mathrm{d}z,\mathrm{d}t) \\
&= -\left\{\begin{array}{l}-w_1(Q_t-q_t)-(\rho-M_1+\delta)p_t^{(Q)}+p_t^{(V)}+ \\ \int_0^\infty\left[\theta^{(Q)}(t,Q,z)+\theta^{(q)}(t,Q,z)+\theta^{(V)}(t,Q,z)\right]\nu(\mathrm{d}z)\end{array}\right\}\mathrm{d}t+\int_0^\infty\int_0^{Q_{t-}}\theta^{(Q)}(t-,u,z)\tilde{N}(\mathrm{d}u,\mathrm{d}z,\mathrm{d}t)
\end{aligned}
,\tag{20}
$$

$$
\begin{aligned}
\mathrm{d}p_t^{(q)} &= -\frac{\partial H}{\partial q}\mathrm{d}t+\int_0^\infty\int_0^{Q_{t-}}\theta^{(q)}(t-,u,z)\tilde{N}(\mathrm{d}u,\mathrm{d}z,\mathrm{d}t) \\
&= -\left\{-w_1(q_t-Q_t)-\frac{\mathrm{d}j(q_t)}{\mathrm{d}q}-\delta p_t^{(q)}-p_t^{(V)}\right\}\mathrm{d}t+\int_0^\infty\int_0^{Q_{t-}}\theta^{(q)}(t-,u,z)\tilde{N}(\mathrm{d}u,\mathrm{d}z,\mathrm{d}t)
\end{aligned}
,\tag{21}
$$

$$
\begin{aligned}
\mathrm{d}p_t^{(V)} &= -\frac{\partial H}{\partial V}\mathrm{d}t+\int_0^\infty\int_0^{Q_{t-}}\theta^{(V)}(t-,u,z)\tilde{N}(\mathrm{d}u,\mathrm{d}z,\mathrm{d}t) \\
&= -\left\{-w_2\left(V_t-\hat{V}_t\right)-\delta p_t^{(V)}\right\}\mathrm{d}t+\int_0^\infty\int_0^{Q_{t-}}\theta^{(V)}(t-,u,z)\tilde{N}(\mathrm{d}u,\mathrm{d}z,\mathrm{d}t)
\end{aligned}
\tag{22}
$$

for $t<T$ and the terminal condition $p_T^{(q)}=p_T^{(Q)}=p_T^{(V)}=0$. We also consider a formal Markovian counterpart by taking the conditional expectation $\mathbb{E}\left[\cdot|\mathcal{F}_t\right]$ of (20)-(22):

$$\mathrm{d}p_t^{(Q)} = \mathbb{E}\left[\left\{\begin{array}{l} w_1\left(Q_t - q_t\right) + \left(\rho - M_1 + \delta\right) p_t^{(Q)} - p_t^{(V)} \\ -\int_0^\infty \left[\theta^{(Q)}\left(t,Q,z\right) + \theta^{(q)}\left(t,Q,z\right) + \theta^{(V)}\left(t,Q,z\right)\right] v\left(\mathrm{d}z\right) \end{array}\right\} \mathrm{d}t \middle| \mathcal{F}_t \right], \tag{23}$$

$$\mathrm{d}p_t^{(q)} = \mathbb{E}\left[\left\{ w_1\left(q_t - Q_t\right) + \frac{\mathrm{d}j\left(q_t\right)}{\mathrm{d}q} + \delta p_t^{(q)} + p_t^{(V)} \right\} \mathrm{d}t \middle| \mathcal{F}_t \right], \tag{24}$$

$$\mathrm{d}p_t^{(V)} = \mathbb{E}\left[\left\{ w_2\left(V_t - \hat{V}_t\right) + \delta p_t^{(V)} \right\} \mathrm{d}t \middle| \mathcal{F}_t \right]. \tag{25}$$

An advantage of using (23)-(25) instead of (20)-(22) is that we do not have to directly handle the integral terms having the random measures. A disadvantage is that the conditional expectations must be evaluated.

By invoking (16), set the candidate optimal control

$$\tilde{a}_t = \max\left\{-\bar{a}, \min\left\{\bar{a}, \frac{p_t^{(q)}}{w_3}\right\}\right\}. \tag{26}$$

We expect that solving the initial value problem of the forward SDEs (1)-(3) and the terminal value problem of the BSDEs (20)-(22) (or (23)-(25)) with $a = \tilde{a}$ of (26) gives the optimal control.

### 2.4.2 Constrained case

We consider FBSDEs of the constrained case heuristically. In this case, it seems difficult to apply a conventional stochastic maximum principle in the earlier section because of the state-constraint that induces a boundary discontinuity of the control [42]. Instead, as in Yoshioka [60], we couple the constrained state dynamics with the BSDE of that obtained in the previous sub-section ((23)-(26)) with an additional assumption (8) to find sub-optimal controls with which the constraint is still activated. Namely, we only change the forward system dynamics and the admissible set of controls. The equation (26) to determine the optimal control is then modified as

$$\hat{a}_t = \underset{a \in A(q_t)}{\arg\max}\, H = \begin{cases} \max\left\{0, \min\left\{\bar{a}, \frac{p_t^{(q)}}{w_3}\right\}\right\} & \left(q_t = 0\right) \\ \max\left\{-\bar{a}, \min\left\{\bar{a}, \frac{p_t^{(q)}}{w_3}\right\}\right\} & \left(0 < q_t < \bar{q}\right) \\ \max\left\{-\bar{a}, \min\left\{0, \frac{p_t^{(q)}}{w_3}\right\}\right\} & \left(q_t = \bar{q}\right) \end{cases}. \tag{27}$$

For the constraint $V_t \in \Omega_V$, we use the singular variables to constrain the dynamics as shown in the second line of (3). Numerically, it is considered as a projection as explained in **Section 4**.

## 3. Mathematical analysis

### 3.1 Sufficient maximum principle

We show that solving the FBSDEs is sufficient for finding the optimal control under the unconstrained case

$\mathcal{A} = \mathcal{A}_1$. **Proposition 1** is the sufficient maximum principle for our model. This proposition applies to cases with $j(q) \neq 0$ and $\bar{a} < +\infty$ where the problem is not necessarily LQ. This is the first sufficient stochastic maximum principle for FBSDEs of reservoir systems with CBI processes.

***Proposition 1*** *Assume that there exists an $\mathcal{F}$-adapted integrable solution $\left(\tilde{Q}_t, \tilde{q}_t, \tilde{V}_t, \tilde{p}_t^{(Q)}, \tilde{p}_t^{(q)}, \tilde{p}_t^{(V)}, \tilde{\theta}^{(Q)}(t,\cdot,\cdot), \tilde{\theta}^{(q)}(t,\cdot,\cdot), \tilde{\theta}^{(V)}(t,\cdot,\cdot)\right)_{t\geq 0}$ to (20)-(22) with (26) such that*

$$\mathbb{E}\left[\int_0^T \left(X_s^2 + |X_s Y_s| + Y_s^2\right) \mathrm{d}s\right] < +\infty, \quad \mathbb{E}\left[\int_0^T \int_0^{\tilde{Q}_t} \int_0^\infty \tilde{\theta}^2(s,u,z) \nu(\mathrm{d}z) \mathrm{d}u \mathrm{d}s\right] < +\infty, \tag{28}$$

*where $X$ represents one of $\left(\tilde{p}^{(Q)}, \tilde{p}^{(q)}, \tilde{p}^{(V)}\right)$, $Y$ represents one of $(Q,q,V)$ for arbitrary $a \in \mathcal{A}_1$, and $\tilde{\theta}$* represents *one of $\left(\tilde{\theta}^{(Q)}, \tilde{\theta}^{(q)}, \tilde{\theta}^{(V)}\right)$. Then, the control of (26) is optimal.*

***Remark 2*** The integrability condition is assumed to guarantee boundedness of the difference of two objective functionals in the proof (See, also the proof of Theorem 3.4 of Hess [31]). The non-negativity assumption of the adjoint variables multiplied by the martingale terms ( $\theta^{(Q)}, \theta^{(q)}, \theta^{(V)}$ in our case) was used in the same literature but is unnecessary here because the process $(Q_t)_{t\geq 0}$ is not controlled. See, (70) in **Appendix C**. Instead, our system is higher dimensional.

### 3.2 An exact solution

We show that the proposed model is solvable analytically in the LQ case.

***Proposition 2*** *In the unconstrained case $\mathcal{A} = \mathcal{A}_1$ with $j(q) \equiv 0$ and $\bar{a} = +\infty$, for $t < T$ we have the following representation of adjoint variables*

$$p_t^{(Q)} = AQ_t + Bq_t + CV_t + D, \tag{29}$$

$$p_t^{(q)} = EQ_t + Fq_t + GV_t + I, \tag{30}$$

$$p_t^{(V)} = JQ_t + Kq_t + LV_t + O, \tag{31}$$

$$\theta^{(Q)}(t,u,z) = Az, \quad \theta^{(q)}(t,u,z) = Ez, \quad \theta^{(V)}(t,u,z) = Jz, \tag{32}$$

*where the coefficients $A, B, C, D, E, F, G, I, J, K, L, O$ are time-dependent and governed by the following system of Riccati equations subject to a homogenous terminal condition at $t = T$ :*

$$A' = w_1 + \left(2\rho(1 - M_1) + \delta\right)A - 2C - \frac{1}{w_3}B^2, \tag{33}$$

$$B' = -w_1 + \left(\rho(1 - M_1) + \delta\right)B + C - G - \frac{1}{w_3}BF, \tag{34}$$

$$C' = \left(\rho\left(1 - M_1\right) + \delta\right)C - L - \frac{1}{w_3}BG, \tag{35}$$

$$D' = -\rho \underline{Q} A + \left(\rho\left(1 - M_1\right) + \delta\right)D - O - \frac{1}{w_3}BI - \rho M_1\left(A + B + C\right), \tag{36}$$

$$F' = w_1 + \delta F + 2G - \frac{1}{w_3}F^2, \tag{37}$$

$$G' = \delta G + L - \frac{1}{w_3}FG, \tag{38}$$

$$I' = -\rho \underline{Q} B + \delta I + O - \frac{1}{w_3}FI, \tag{39}$$

$$L' = w_2 + \delta L - \frac{1}{w_3}G^2, \tag{40}$$

$$O' = -\rho \underline{Q} C - w_2 \hat{V} + \delta O - \frac{1}{w_3}GI. \tag{41}$$

*with* $E = B$, $J = C$, $K = G$.

The Riccati equations can be solved numerically using a conventional ODE solver like a classical explicit Euler method. We use this numerical method in this paper. The system admits a unique local solution because its coefficients are at most quadratic and hence are locally Lipschitz continuous.

***Remark 3*** The LQ case complies with **Proposition 1** at least for not large $T > 0$ because the Riccati equations have quadratic coefficients. In addition, by the symmetry $E = B$, $J = C$, $K = G$, the coefficients $E, F, G, I, J, K, L, O$ are found without using $A, B, C, D$. This corresponds to the fact that the optimal control is independent from $p^{(Q)}$ as also discussed in the next section.

## 4. Numerical computation

### 4.1 Numerical scheme

We introduce a numerical scheme for FBSDEs based on a least-squares Monte-Carlo method and a Picard-type iteration. The strength of the proposed approach is that the optimal controls can be found without directly optimizing the objective functional.

Our scheme handles all the forward dynamics and the adjoint ones (24)-(25), but not (23) because it is unnecessary. In fact, the optimal control depends on $p^{(q)}$, and $p^{(q)}$ depends on $p^{(V)}$ but not on $p^{(Q)}$. In this view, it is sufficient to solve the forward dynamics (1)-(3), the adjoint equations (24)-(25), and (16) or (27). This connection among the adjoint variables is owing to the uncontrolled nature of the inflow $Q$. Disregarding the adjoint equation of $p^{(Q)}$ is advantageous from the two numerical viewpoints. Firstly, we can simply save computational costs. Secondly, we can avoid approximating the three maps $\theta^{(Q)}, \theta^{(q)}, \theta^{(V)}$ modulating the Martingale parts of the BSDEs. Similar maps commonly arise in BSDEs driven by Lévy jumps, but their approximation seems not to be straightforward [69].

### 4.1.1 Forward SDEs

Discretization of the forward SDEs is based on a classical explicit Euler-Maruyama type scheme. The time grid is prepared as $t_i = ih$ ($i = 0,1,2,...,n$) with $n \in \mathbb{N}$ and the time increment $h = n^{-1}T$. The discretized state variable at time $t = t_i$ is denoted using the indicator $[i]$ like $Q[i]$. Assume that a candidate of a temporally discretized Markov control $a^*[i]$ ($i = 0,1,2,...,n-1$) is given. Its computation method will be presented in the next sub-section.

For the unconstrained case, using the initial conditions $Q[0] = Q_0$, $q[0] = q_0$, $V[0] = V_0$, the forward SDEs for $i \geq 1$ are discretized as follows:

$$Q[i+1] = Q[i] + \rho h\left(\underline{Q} - Q[i]\right) + Z[i], \tag{42}$$

$$q[i+1] = q[i] + a[i]h, \tag{43}$$

$$V[i+1] = V[i] + \left(Q[i] - q[i]\right)h. \tag{44}$$

Here, the variable $Z[i]$ to approximate $\int_{t_i}^{t_{i+1}} \int_0^{\infty} \int_0^{Q_{t-}} zN(\mathrm{d}u, \mathrm{d}z, \mathrm{d}t)$ is generated using an acceptance rejection method [70]. An extension here is that the jump measure is proportional to $Q[i]$ as $Z[i] = TS\left(\alpha, aQ[i]h, b\right)$, where the map $TS(\cdot,\cdot,\cdot)$ to sample tempered stable variables is explained in Algorithm 0 of Kawai and Masuda [70] and is not repeated here. In the constrained case, we apply the projection [71] to constrain the water volume:

$$V[i+1] = \max\left\{0, \min\left\{\bar{V}, V[i] + \left(Q[i] - q[i]\right)h\right\}\right\}. \tag{45}$$

### 4.1.2 Temporal discretization of backward SDEs

The BSDEs are discretized in a time-backward manner using the terminal conditions $p^{(q)}[n] \equiv p^{(V)}[n] \equiv 0$. According to (24)-(25), an adjoint variables at each time $t$ can be formally understood as a function of $(t, Q_t, q_t, V_t)$. Therefore, discretized adjoint variables at each $t = t_i$ are considered as 3-D mappings from $(0,+\infty) \times \mathbb{R}^2$ to $\mathbb{R}$ in the unconstrained case and from $(0,+\infty) \times \Omega_q \times \Omega_V$ to $\mathbb{R}$ in the constrained case. We use the notations like $p^{(q)}[i] \equiv p^{(q)}[i](Q,q,V)$. By a fully-implicit discretization, for $0 \leq i \leq n-1$, we discretize (24) and (25) as

$$-p^{(q)}[i](Q,q,V) = \mathbb{E}\left[-p^{(q)}[i+1] + \left\{w_1\left(q[i] - Q[i]\right) + \frac{\mathrm{d}j\left(q[i]\right)}{\mathrm{d}q} + \delta p^{(q)}[i] + p^{(V)}[i]\right\}h \middle| \mathcal{F}_{t_i}\right] \tag{46}$$

and

$$-p^{(V)}[i](Q,q,V)=\mathbb{E}\left[-p^{(V)}[i+1]+\left\{w_2\left(V[i]-\hat{V}_{t_i}\right)+\delta p^{(V)}[i]\right\}h\middle|\mathcal{F}_{t_i}\right], \quad (47)$$

respectively. By a tower property (47) becomes

$$p^{(V)}[i](Q,q,V)=\frac{1}{1+\delta h}\mathbb{E}\left[p^{(V)}[i+1]-\left\{w_2\left(V[i]-\hat{V}_{t_i}\right)\right\}h\middle|\mathcal{F}_{t_i}\right]. \quad (48)$$

Similarly, by (48), (46) is rewritten as

$$p^{(q)}[i](Q,q,V)=\frac{1}{1+\delta h}\mathbb{E}\left[\begin{array}{l}p^{(q)}[i+1]-\left\{w_1\left(q[i]-Q[i]\right)+\frac{\mathrm{d}j\left(q[i]\right)}{\mathrm{d}q}\right\}h\\-\frac{h}{1+\delta h}\left(p^{(V)}[i+1]-\left\{w_2\left(V[i]-\hat{V}_{t_i}\right)\right\}h\right)\end{array}\middle|\mathcal{F}_{t_i}\right]. \quad (49)$$

### 4.1.3 Stochastic grid bundling

We employ a lest-squares Monte-Carlo method to approximate each conditional expectation in (48)-(49) at each time step. We also use the stochastic grid bundling [63] with a smooth monitor function to determine spatial grid at each time step. In our computation, a conditional expectation of the form $\mathbb{E}\left[\cdot\middle|\mathcal{F}_{t_i}\right]$ is considered as a function of state variables $\left(Q_{t_i},q_{t_i},V_{t_i}\right)$ at time $t=t_i$. A naïve approach is to regress a conditional expectation using global polynomials [40], but this approach is subject to high variance when using high-order polynomials. Locally and regularly splitting the state space into small grids and regressing the conditional expectation in each grid is another possibility but is subject to huge computational costs. For an extensive discussion on different basis, we refer to Bender and Steiner [72].

The stochastic grid bundling was proposed in Jain and Oosterlee [73] to construct a space grid at each time step so that the locations and sizes of bundles as components of the grid are determined adaptively. The stochastic grid bundling of Chau and Oosterlee [63] is based on this method enhanced with an efficient monitor function so that the equal number of sample paths fall on each bundle.

Assume that we have $S\in\mathbb{N}$ sample paths and $S_{\mathrm{bundle}}$ bundles, and further that $n_{\mathrm{bundle}}=S/S_{\mathrm{bundle}}\in\mathbb{N}$. We show how to generate a grid at time $t=t_i$. Each path is ranked by an index $\psi=\psi\left(Q_{t_i},q_{t_i},V_{t_i}\right)$. Its value of the $p$ th path is denoted as $\psi_p$ ($1\leq p\leq S$). The simplest example is $\psi\left(Q_{t_i},q_{t_i},V_{t_i}\right)=Q_{t_i}$ used later. We then have the collection $\left\{\psi_p\right\}_{1\leq p\leq S}$. There is a surjective mapping $\Xi(\cdot)$ from $\{1,2,3,...,S\}$ to $\{1,2,3,...,S\}$ such that

$$\psi_{\Xi(1)}\leq\psi_{\Xi(2)}\leq...\leq\psi_{\Xi(S-1)}\leq\psi_{\Xi(S)}. \quad (50)$$

One map $\Xi$ satisfying (50) is computed by a common sorting algorithm. We use the quick sort [74]. After sorting the paths as in (50), we divide all paths into $n_{\mathrm{bundle}}$ bundles. The $i_{\mathrm{bundle}}$ th bundle ($1\leq i_{\mathrm{bundle}}\leq n_{\mathrm{bundle}}$), denoted as $G\left(i_{\mathrm{bundle}}\right)$, is a collection of natural numbers: $G\left(i_{\mathrm{bundle}}\right)=\left\{i_{\mathrm{bundle}}\,p\right\}_{1\leq p\leq S_{\mathrm{bundle}}}$.

We define cells as physical counterparts of bundles. Set the cell boundaries as

$$\bar{B}(i_{\text{bundle}}) = \underline{B}(i_{\text{bundle}}+1) = \frac{\psi_{\Xi(i_{\text{bundle}}p)} + \psi_{\Xi(i_{\text{bundle}}p+1)}}{2} \quad (1 \le i_{\text{bundle}} \le S_{\text{bundle}} - 1) \tag{51}$$

with $\underline{B}(1) = -\infty$ and $\bar{B}(S_{\text{bundle}}) = \infty$. We then consider $i_{\text{bundle}}$ th cell $C(i_{\text{bundle}})$ as the interval $C(i_{\text{bundle}}) = \left(\underline{B}(i_{\text{bundle}}), \bar{B}(i_{\text{bundle}})\right]$ ( $2 \le i_{\text{bundle}} \le S_{\text{bundle}} - 1$ ) with $C(1) = \left(-\infty, \bar{B}(1)\right]$ and $C(S_{\text{bundle}}) = \left(\underline{B}(S_{\text{bundle}}), +\infty\right)$. The expectation $\mathbb{E}\left[\cdot \middle| \mathcal{F}_{t_i}\right]$ is then evaluated for each bundle or equivalently in each cell. With the grid bundling, we obtain the approximation $E_{i,i_{\text{bundle}}}(X) = E_{i,i_{\text{bundle}}}(X)(Q,q,V)$ of $\mathbb{E}\left[X \middle| \mathcal{F}_{t_i}\right]$ for each random variable $X$ as a function of the state variables at $t_i$:

$$E_i(X)(Q,q,V) = \sum_{i_{\text{bundle}}=1}^{n_{\text{bundle}}} \chi_{\{\psi(Q,q,V) \in C(i_{\text{bundle}})\}} E_{i,i_{\text{bundle}}}(X)(Q,q,V). \tag{52}$$

The approximations of the adjoint variable $p^{(q)}$ ( $p^{(V)}$ ) is then obtained by substituting $p^{(q)}$ ( $p^{(V)}$ ) into (52). Theoretically, the total number of bundles $S_{\text{bundle}}$ can be different at each time step. We fix $S_{\text{bundle}} = 1$ at $t_0 = 0$ because we use a deterministic initial value.

***Remark 3*** Our discretization is of a regress now type that is subject to statistical errors in evaluating the conditional expectations $\mathbb{E}\left[p^{(q)}[i+1] \middle| \mathcal{F}_{t_i}\right]$ and $\mathbb{E}\left[p^{(V)}[i+1] \middle| \mathcal{F}_{t_i}\right]$ because of using $\left(Q_{t_i}, q_{t_i}, V_{t_i}\right)$ to regress them. Employing a regress later type method that regresses the conditional expectations using $\left(Q_{t_{i+1}}, q_{t_{i+1}}, V_{t_{i+1}}\right)$ can eliminate the statistical error but requires an analytical representation formula of them [63]. This is possible for Brownian noises but seems to be difficult for CBI cases.

#### 4.1.4 Relaxed Picard iteration

We implement a Picard-type algorithm that numerically solves the forward and backward parts in an alternating manner (Section 4.2 of Chassagneux et al. [40]). The algorithm is the same for both the constrained and unconstrained cases except for the treatment of the state constraint.

**(Pseudo-code)**

1. Set coefficients and their parameter values.
2. Set the parameter values of numerical discretization ( $n, S, S_{\text{bundle}}$ ) and basis.
3. Prepare an initial guess $p^{(q)}(0)$ of $E_i\left(p^{(q)}\right)(q,Q,V)$ $(1 \le i \le n-1)$ at the 0 th Picard iteration. Similarly, $E_i\left(p^{(q)}\right)(q,Q,V)$ $(1 \le i \le n-1)$ at $i_{\text{Picard}}$ th iteration is denoted as $p^{(q)}(i_{\text{Picard}})$.
4. Set $i_{\text{Picard}} = 1$.
5. Simulate $S$ independent paths of $\left(Q_{t_i}, q_{t_i}, V_{t_i}\right)$ $(1 \le i \le n-1)$ with numerically evaluated $\hat{a}$.

6. Compute $E_i\left(p^{(q)}\right)(Q,q,V)$ and $E_i\left(p^{(V)}\right)(Q,q,V)$ $(1 \le i \le n-1)$ backward in time.
7. Check a convergence criterion. If it is satisfied, then stop the algorithm and output the latest computed values of the forward and backward parts. If it is not, then $i_{\text{Picard}} \to i_{\text{Picard}} + 1$ and go to Step 5.

As the convergence criterion at Step 7, we choose the following one because we use a deterministic initial condition $(Q_0, q_0, V_0)$ in each numerical simulation:

$$\left| p^{(q)}\left(i_{\text{picard}}\right) - p^{(q)}\left(i_{\text{picard}} - 1\right) \right| \le \varepsilon \quad \text{at the initial time } t_0 = 0, \tag{53}$$

where $\varepsilon$ should be a small value to judge convergence. We choose $\varepsilon = 10^{-6}$.

Numerical evaluation of $\hat{a}$ is crucial for convergence. As Kerimkulov et al. [64] discussed, for the unconstrained case, the iteration with the following naïve choice did not converge in our problems

$$\hat{a}_i\left(i_{\text{Picard}}\right) = \max\left\{-\bar{a}, \min\left\{\bar{a}, \frac{E_i\left(p^{(q)}\left(i_{\text{Picard}}\right)\right)}{w_3}\right\}\right\}, \tag{54}$$

while the following one with a sufficiently large positive parameter $\rho > 0$ works in our case:

$$\hat{a}_i\left(i_{\text{Picard}}\right) = \max\left\{-\bar{a}, \min\left\{\bar{a}, \frac{\rho}{w_3+\rho}\hat{a}_i\left(i_{\text{Picard}} - 1\right) + \left(1 - \frac{\rho}{w_3+\rho}\right)\frac{E_i\left(p^{(q)}\left(i_{\text{Picard}}\right)\right)}{w_3}\right\}\right\}, \tag{55}$$

which is a relaxation approximation. The quantity $\frac{\rho}{w_3+\rho} \in (0.1)$ can be seen as a relaxation factor weighting more on the previous value by choosing it a s larger value. The same technique is applied to the constrained case. We specify $\rho$ so that $\frac{\rho}{w_3+\rho} = 0.5$.

### 4.2 Study site

The study site is the Obara Dam and Sakura-Orochi Reservoir system (OSO system in short) located at mid-stream of Hii River, Japan. Its capacity is $\bar{V} = 6 \times 10^7$ (m$^3$). This system is operated by Ministry of Land, Infrastructure, Transport and Tourism of Japan, and has been a study site of a series of recent research on management of water environment and ecosystems, especially bloom of benthic algae in the dam-downstream river [75], fishery resource management [76], and water quality analysis [23]. The main purposes of OSO system are water resources supply for irrigation and drinking, flood mitigation against its downstream area, and recreations including sightseeing and boat race in the reservoir. The OSO system has also been studied from a viewpoint of dam and reservoir systems management using a Markov chain [42].

Operation data of the OSO system at every hour is available since April 1 in 2016 [77]. **Figure 2** presents the hourly inflow discharge of the system in the fiscal years 2016 to 2019. Each fiscal year is from April 1 to the coming March 31 in Japan. The inflow discharge has a stochastic and clustering nature corresponding to rainfalls in rainy season (May to July), typhoon events (September to October), and snow melting (March to April). We use the four-year data from April 1 in 2016 to March 31 in 2020 because a hydropower station located upstream of the OSO system, which was abstracting the water from the river, has been idling since May in 2021. Studying the impacts of idling the hydropower station on its downstream river environment is beyond the scope of this paper but can be analyzed in future.

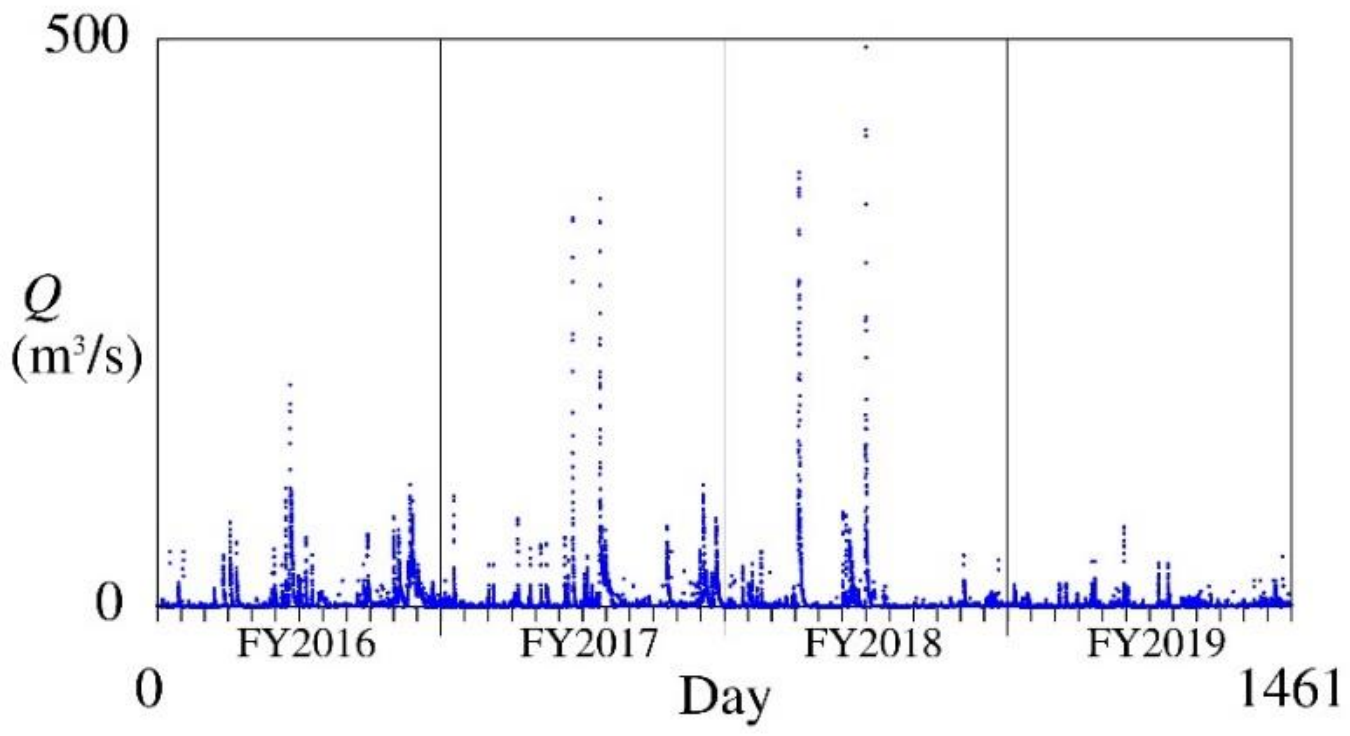

**Figure 2.** Hourly inflow discharge of the OSO system from the fiscal years (FYs) 2016 to 2019. Each FY is from April 1 to the coming March 31 in Japan.

### 4.3 Identification of the model and specification of parameter values

We identify the model parameters of the SDE (1): $\underline{Q}$, $a$, $b$, $\alpha$, and $\rho$. Set $\underline{Q} = 0.50$ (m$^3$/s). The remaining parameters of the dynamics are $a$, $b$, and $\alpha$ of $v$. We use the method of matching moments of $Q_t$ based on average (Ave), standard deviation (Sta), skewness (Ske), and kurtosis (Kur) for a stationary state. Kur is set to be 0 for Gaussian distributions. The three parameters are optimized using a classical nonlinear least-squares method minimizing the sum of the relative errors between empirical and modelled values of the four moments. The analytical procedures to calculate the moments are provided in **Appendix B**. We get $\alpha = 0.923$ (-), $a = 0.0493$ ( $\mathrm{s}^{(\alpha-1)} / \mathrm{m}^{3(\alpha-1)}$ ), and $b = 0.007$ (s/m$^3$). The empirical and modelled moments are comparted in **Table 1**. A good agreement between them is observed especially for Sta and Kur. The relative error of Ave, which is the most important statistics, is 0.17%. We have $1 - M_1 = 0.095$ and hence assumption (5) is satisfied. Secondly, we have estimated the autocorrelation function of the inflow discharge as $e^{-\rho_c \tau}$ with $\rho_c = 0.028$ (1/h) where $\tau$ is the time lag [23]. Then, as in the Hawkes [78], we get $\rho_c = (1 - M_1)\rho$, from which we obtain $\rho$ as $\rho_c (1 - M_1)^{-1} = 0.295$ (1/h).

We use the discount rate $\delta = 1/12$ (day$^{-1}$) or equivalently 2 (h$^{-1}$). This is much larger than the inverse of the terminal time $T^{-1}$, meaning that the timescale of decision-making is shorter than that of the problem horizon. We have the two reasons. Firstly, it is natural to assume that the decision-maker, the operator of the dam and reservoir system, make decisions with a timescale not larger than a day. In fact, tracking the inflow discharge to modulate the outflow discharge requires a timescale smaller than the correlation timescale of the inflow, which in our case is $\rho^{-1} = 37$ (h). Secondly, even for the LQ case, specifying a smaller $\delta$ such as $\delta = 1/24$ (day$^{-1}$) does not give converged numerical solutions with the relaxed Picard iteration. This is a technical limitation because intuitively the BSDE should be solvable for all $\delta > 0$. Using a stronger relaxation in the iteration procedure may resolve this issue, while the numerical scheme becomes critically inefficient at the same time. Development of a scheme that can manage small $\delta$ in an efficient and stable manner is necessary to overcome this issue, which will be addressed in our future research. We emphasize here that the presented numerical approach is a first attempt toward modeling and control of dam and reservoir systems based on FBSDEs and a CBI process.

In the numerical computation, the reservoir water volume $V_t$ is scaled as $V_t \to \left(3600/\bar{V}\right) \cdot V_t$, where the factor 3600 comes from the fact that the discharges are usually evaluated in the unit (m$^3$/s) while our interest is numerical computation in a timescale not shorter than hours. The division by $\bar{V}$ is for normalizing the water volume by its maximum. The range of the $V_t$ is then rescaled as $\left[0, \bar{V}\right]$ with the (non-dimensionalized) maximum water volume $\bar{V} = 16666,7$.

The other model parameters of the objective functional are specified as well. We choose $w_1 = 1$, $w_2 = 4/\bar{V}$, $w_3 = 2000$, $\hat{V} = \bar{V}/2$ unless otherwise specified. Furthermore, set $j(q) \equiv 0$ for the LQ case while use $j(q) = \frac{w_4}{2} \max\left\{\underline{q} - q, 0\right\}^2$ for the non-LQ case ( $w_4 = 1$ and $\underline{q} = 5.0$ (m$^3$/s)). The dimensions of the weighting factors are determined so that the integrand of the objective functional (10) is dimensionless. This non-dimensionalization is not problematic because what is important here is not the absolute values of each of the terms of the integrand but their relative magnitudes. Finally, for the LQ case set $T = 60$ (day) and for the non-LQ case $T = 360$ (day). Set $\bar{a} = 100$ (m$^3$/s/h) in the constrained case. We fix the initial condition $\left(Q_0, q_0, V_0\right) = \left(0.5 \text{ m}^3/\text{s}, 0.5 \text{ m}^3/\text{s}, \bar{V}/2\right)$ in the computation below.

**Table 1.** Comparison of empirical and modelled statistical moments.

| | Ave (m$^3$/s) | Sta (m$^3$/s) | Ske (-) | Kur (-) |
|---|---|---|---|---|
| Empirical | 5.290 | 16.58 | 12.84 | 258.1 |
| Modelled | 5.281 | 16.60 | 12.45 | 258.1 |
| Relative error (%) | 0.17 | 0.00 | 2.98 | 0.00 |

### 4.4 Basis functions and regularization

The sets of basis considered in our computation are $\mathcal{S}_{\text{LQ}} = \{1, Q_t, q_t, V_t\}$, $\mathcal{S}_{\text{NLQ1}} = \left\{\mathcal{S}_{\text{LQ}}, \max\left\{\underline{q} - q_t, 0\right\}\right\}$, and $\mathcal{S}_{\text{NLQ2}} = \left\{\mathcal{S}_{\text{NLQ1}}, Q_t \max\left\{\underline{q} - q_t, 0\right\}, V_t \max\left\{\underline{q} - q_t, 0\right\}\right\}$. Here, $\mathcal{S}_a = \left\{\mathcal{S}_b, \mathcal{S}_c\right\}$ means that the set $\mathcal{S}_a$ is a collection of all the elements of mutually independent sets $\mathcal{S}_b$ and $\mathcal{S}_c$. The first set $\mathcal{S}_{\text{LQ}}$ corresponds to the exact basis of the LQ case (**Proposition 2**). The second set $\mathcal{S}_{\text{NLQ1}}$ and third set $\mathcal{S}_{\text{NLQ2}}$ are used for the non-LQ cases with the aforementioned $j(q)$. The choice of the basic functions based on $\max\left\{\underline{q} - q, 0\right\}$ is motivated by the intuition based on the simpler problem [42, 24] that the value functions in the non-LQ cases would have an inflection point at $q = \underline{q}$, at which it is not necessarily smooth.

The least-squares Monte-Carlo method requires inverting a correlation matrix to find the coefficients to be multiplied by the basis. This matrix becomes ill-conditioned whose inversion should be carried out with a regularization method. We use the ridge regularization [79] with the regularization parameter $\lambda > 0$, with which the correlation matrix becomes invertible. We use $\lambda = 10^{-8}$ in **Section 4.5** for LQ cases and $\lambda = 10^{-7}$ in **Section 4.6** for non-LQ cases. We examine several values of $\lambda$ in the numerical experiment for the LQ case to imply that there would be an optimal $\lambda$ minimizing the computational error for each computational condition. Simply setting $\lambda = 0$ did not work in the proposed model. A conjugate gradient method is employed to invert correlation matrices.

### 4.5 LQ case: comparison with the exact solution

We analyze computational performance of the numerical scheme against the unconstrained case. We fix a relationship between the time increment $h$ (h) and the total number $S$ of sample paths as $S = 10000h^{-0.5}$ considering the formal theoretical result that classical Euler-Maruyama schemes for SDEs has the error $O(h)$ in the weak sense and that the and the fact that the Monte-Carlo simulation induces the statistical error $O(S^{-0.5})$ (Glasserman and Merener [80] Section 3.5 of Seydel [81]). We examine $(h, S)$ of $(2.0, 2500)$, $(1.0, 10000)$, $(0.5, 40000)$, and $(0.25, 160000)$. The terminal time is set as $T = 120$ (day). We use the basis $\mathcal{S}_{\mathrm{LQ}}$, which is the exact choice according to **Proposition 2**. We thus focus on performance of the scheme without computational errors induced by basis. Due to the choice of the basis, we do not active the grid bundling in the computation here ($n_{\mathrm{bundle}} = 1$). By preliminary numerical experiments not presented here, we found that the parameter $w_3$ for penalizing fast variations of the outflow discharge is a parameter affecting both the controlled outflow discharge and water volume. We therefore examine the four values of $w_3$ as a parameter study: 1000, 2000, 3000, and 4000.

**Figure 3** compares the reference and numerical solutions, demonstrating that numerical solutions indeed converge with smaller errors for finer resolution. **Tables 2-3** show $l^1$ and $l^2$ errors of the coefficient $E = E_t$ of the adjoint variable $p^{(q)}$ with different values of $h$ and $w_3$ over the time interval $(0, T]$. We have chosen this coefficient as it causally links the inflow discharge with the control. The $l^p$ error ($p = 1, 2$) is calculated in a standard way as $e_n^p = \left( \frac{1}{n} \sum_{i=0}^{i=n} \left| E_{t_i,\mathrm{num}} - E_{t_i,\mathrm{ref}} \right|^p \right)^{1/p}$, where the subscripts "num" and "ref" represent numerical and reference solutions, respectively. Yoshioka [60] employed a similar error criterion for least-squares Monte-Carlo discretization of FBSDEs driven by a tempered stable Ornstein-Uhlenbeck (TSOU) process.

The computational results suggest that the convergence rates averaged for different computational resolutions are smaller than 1 for all the cases. They are near 0.5 for the $l^2$ error and near 0.86 for the $l^1$ error. These convergence rates (CRs, calculated as $\log_2\left(e_n^p / e_{2n}^p\right)$) are worse than nearly 1 observed for FBSDEs with a TSOU process [60]. This is possibly because of using a positive regularization parameter $\lambda$ with which numerical solutions do not strictly converge to the exact solution due to regularization errors. In fact, no regularization was necessary in Yoshioka [60]. Another cause of this difference of the convergence rates would be attributed to the fact that the TSOU process is driven by state-independent jumps while the CBI process by state-dependent jumps. The state-dependence prevented us from employing the exact discretization of tempered stable jumps as a low variance numerical method [82]. To date, we did not find any exact discretization methods for CBI processes.

We carried out another set of numerical computation with different values of $\lambda$ using $w_3 = 1000$ and $(h, S) = (0.5, 40000)$. **Table 4** suggests that there is a best $\lambda$ minimizing the errors. Using $\lambda$ around $10^{-7}$ is a reasonable choice in this test case. Another finding is that using the best $\lambda$ improves the computational errors but not their convergence rates. As discussed in Hu and Zastawniak [79] it seems to be difficult to theoretically identify the minimizing $\lambda$ given a computational condition. Numerical computation with the finer resolution suggested that the best $\lambda$ is smaller than $10^{-7}$ (**Table 5**). Therefore, the choice $\lambda = 10^{-7}$ will work well only for the former test case. The finding that smaller $\lambda$ is better for larger number of sample paths is consistent with the existing results [79].

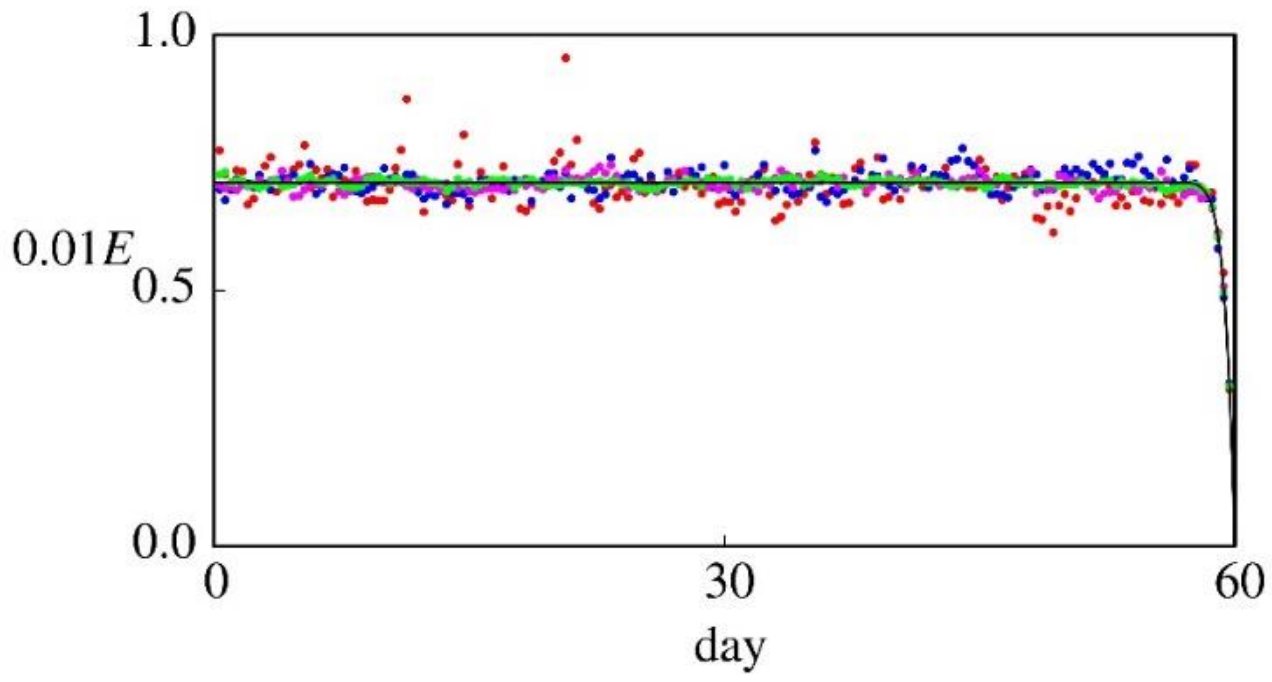

**Figure 3.** The reference solution (Black), numerical solutions with $h = 2$ (Red), $h = 1$ (Blue), $h = 0.5$ (Pink), and $h = 0.25$ (Green) for $w_3 = 3000$.

**Table 2.** $l^1$ error of the coefficient $E = E_t$ of the adjoint variable $p^{(q)}$ with different values of $h$ and $w_3$. We fix $\lambda = 10^{-8}$. The CR averaged for different computational resolutions is also reported.

| $h$ | $w_3 = 1000$ | $w_3 = 2000$ | $w_3 = 3000$ | $w_3 = 4000$ |
|---|---|---|---|---|
| 2.00 | 2.8429 | 3.1707 | 3.3527 | 3.4522 |
| 1.00 | 1.5295 | 1.7433 | 1.8420 | 1.9013 |
| 0.50 | 0.8075 | 0.9112 | 0.9620 | 0.9930 |
| 0.25 | 0.4720 | 0.5303 | 0.5548 | 0.5685 |
| CR | 0.86 | 0.86 | 0.87 | 0.87 |

**Table 3.** $l^2$ error of the coefficient $E = E_t$ of the adjoint variable $p^{(q)}$ with different values of $h$

and $w_3$. We fix $\lambda = 10^{-8}$. The CR averaged for different computational resolutions is also reported.

| $h$ | $w_3 = 1000$ | $w_3 = 2000$ | $w_3 = 3000$ | $w_3 = 4000$ |
|---|---|---|---|---|
| 2.00 | 4.3881 | 3.2624 | 3.4515 | 3.5555 |
| 1.00 | 2.4010 | 2.7777 | 2.9667 | 3.0823 |
| 0.50 | 1.4645 | 1.6903 | 1.8043 | 1.8751 |
| 0.25 | 0.9455 | 1.1009 | 1.1792 | 1.2284 |
| CR | 0.74 | 0.52 | 0.51 | 0.51 |

**Table 4.** $l^1$ and $l^2$ errors of the coefficient $E = E_t$ of the adjoint variable $p^{(q)}$ with different $\lambda$. We fix $h = 0.5$. Both computational errors are minimized at $\lambda = 5.0 \times 10^{-8}$.

| $\lambda$ | $l^1$ error | $l^2$ error |
|---|---|---|
| $1.0 \times 10^{-9}$ | 0.8092 | 1.4656 |
| $5.0 \times 10^{-9}$ | 0.8082 | 1.4656 |
| $1.0 \times 10^{-8}$ | 0.8075 | 1.4645 |
| $5.0 \times 10^{-8}$ | 0.8073 | 1.4607 |
| $1.0 \times 10^{-7}$ | 0.8100 | 1.4612 |
| $5.0 \times 10^{-7}$ | 0.8603 | 1.5078 |
| $1.0 \times 10^{-6}$ | 0.8603 | 1.5078 |

**Table 5.** $l^1$ and $l^2$ errors of the coefficient $E = E_t$ of the adjoint variable $p^{(q)}$ with different $\lambda$. We fix $h = 0.25$. Both computational errors are suggested to be minimized at $\lambda < 1.0 \times 10^{-7}$.

| $\lambda$ | $l^1$ error | $l^2$ error |
|---|---|---|
| $1.0 \times 10^{-10}$ | 0.4687 | 0.9389 |
| $1.0 \times 10^{-9}$ | 0.4679 | 0.9383 |
| $1.0 \times 10^{-8}$ | 0.4719 | 0.9455 |
| $1.0 \times 10^{-7}$ | 0.6068 | 1.0564 |

### 4.6 Non-LQ case

Now the state variables are constrained. We firstly provide demonstrative computational examples with different target water volumes $\left(\hat{V}_t\right)_{t \geq 0}$: **(a)** sine $\hat{V}_t = 0.5\bar{V} + 0.25\bar{V} \sin\left(\frac{2\pi t}{T}\right)$, **(b)** cosine $\hat{V}_t = 0.5\bar{V} + 0.25\bar{V} \cos\left(\frac{2\pi t}{T}\right)$, **(c)** sudden decrease $\hat{V}_t = 0.5\bar{V} - 0.25\bar{V} \chi_{(0.25T, 0.75T]}$, and **(d)** sudden increase $\hat{V}_t = 0.5\bar{V} + 0.25\bar{V} \chi_{(0.75T, 0.25T]}$. The sine case **(a)** is the case where the water volume should be changed following a smooth seasonal target. This is a model case where a gradual regime shift of the water environment of reservoir and its surrounding are preferred. For example, seasonal changes of water levels are natural phenomena in surface water bodies, but sudden artificial drying up of riparian zone surrounding a reservoir may trigger unexpected disasters like landslides [46]. The cosine case **(b)** is analogous to the sine case **(a)** except for that the target $\hat{V}_t$ is not equal to the initial water volume $V_0 = \bar{V}/2$ in the former case. We will see that the proposed numerical method works under this inconsistency. On the contrary, rapid control of the reservoir water volume as in the cases **(c)**-**(d)** has been effectively employed to suppress population increase of invasive fish species as well as to conserve naïve species [19-20]. Both gradual and sudden controls of the reservoir water volume are thus encountered in real applications, which is the reason why we consider both smooth **(a, b)** and sudden **(c, d)** transitions of the target water volume. The case **(d)** is an extreme case in the sense that the two opposite objectives coexist; the outflow discharge should be larger than the threshold $\underline{q}$ while the water volume should track the increased target level. The latter goal forces the decision-maker to choose a smaller discharge, which conflicts with the former one. The proposed optimal control approach would give a compromising solution considering the weighting factors of (11). We employ $(S,h)=(40000, 0.5\ \mathrm{h})$, $w_3 = 2000$, and $n_{\mathrm{bundle}} = 8$ unless otherwise specified.

**Figures 4-7** show a common sample path of the inflow, the controlled sample paths of the outflow and the water volume, and the corresponding probability density of the water volume for the cases **(a)-(d)**. The clustered jumps are simulated accordingly and the outflow discharge is larger than the inflow discharge especially when the latter is smaller than the threshold $\underline{q} = 5.0$ (m$^3$/s). This characteristic was not reproduced under the LQ case as shown in **Appendix B**. The peak discharges are effectively cut off by softly limiting the speed of change of the outflow discharge.

**Figures 4-7** show that the controlled water volume tracks the targeted one with a time lag, which is due not only to softly penalizing deviations but also, we are considering the other control objectives as well. **Figure 8** shows the average and standard deviations of the controlled water volumes with different values of $w_2$, suggesting that both the lag and deviation become smaller as $w_2$ is specified larger. Using a larger $w_2$ would reduce the lag, but the relaxed Picard iteration did not always converge. This is due to the increased growth rate of the driver of the FBSDEs that makes the convergence severer [64]. A theoretical countermeasure to reduce the time lag would be incorporating some anticipative actions to better meet the control objectives. The framework of anticipative control [83] may compensate this issue, but its implementation needs a more versatile algorithm.

Probabilistic nature of the controlled outflow is further analyzed focusing on that near the threshold $\underline{q} = 5.0$ (m$^3$/s). **Figure 9-12** show the computed probability (Pr) of the outflow discharge $q$ (m$^3$/s) for each low discharge levels for the cases **(a)-(d)**. From **Figures 9-10**, the outflow discharge at the low levels is gradually varying in time due to the smooth transitions of $\hat{V}_t$. An exception is the sudden decrees of the probabilities $\Pr(0 \le q < 2)$ and $\Pr(2 \le q < 4)$ near the initial time $t = 0$ during which the outflow discharge is controlled to better meet the threshold $\underline{q}$. The probabilities $\Pr(4 \le q < 6)$, $\Pr(6 \le q < 8)$, and $\Pr(8 \le q < 10)$ are accordingly increasing near $t = 0$. **Figures 11-12** show that the discontinuities of the target water volume $\hat{V}_t$ at $t = 0.25T, 0.75T$ induce sudden transitions of the outflow discharge occurring around these time instances. In particular, the increase (resp., decrease) of the target water volume decreases (resp., increases) the outflow discharge, indicating that the discontinuous transitions of the controlled process can be simulated in a stable manner using the proposed numerical method. Another important finding is that the probabilities of the controlled outflow discharge gradually vary as the time elapses even when $\hat{V}_t$ remains constant. This tendency is more significant for lower levels of $q$. The outflow discharge $q$ should therefore be carefully controlled to meet the operational objective especially when it is lower than the threshold $\underline{q}$.

We finally analyze impacts of basis and bundling on the computed control and dynamics. We focus on the sine case **(a)** and the FBSDEs are computed using the proposed numerical method and the parameter values but with different basis and the total number of bundles $S_{\mathrm{bundle}}$. An emphasis is put on the controlled water volume with which differences among the different computational conditions can be effectively analyzed. **Figures 13-14** compare the averages and standard deviations of the controlled water volumes for different basis and grid bundling, respectively. **Figure 13** shows that the controlled water volumes with the basis sets $\mathcal{S}_{\mathrm{LQ}}$ and $\mathcal{S}_{\mathrm{NLQ2}}$ are comparable both in the average and standard deviation with a larger standard deviation in the latter. The numerical solution with $\mathcal{S}_{\mathrm{NLQ1}}$ is clearly different from the others in the latter half of the computational period. Similar basis works without convergence problems for simpler non-LQ systems having a liner inflow SDE with the computational period of 120 (day) [60]. This peculiar behavior of the numerical solution with $\mathcal{S}_{\mathrm{NLQ1}}$ is considered due to the longer computational period (360 (day)) in the present case where the convergence condition of the relaxed Picard iteration (small $T$ with sufficiently regular coefficients) may not be satisfied numerically [64]. We did not find similar peculiar behavior with the basis $\mathcal{S}_{\mathrm{NLQ2}}$. A rigorous mathematical criterion to judge stability and consistency of the relaxed Picard iteration will be necessary to better understand this issue.

**Figure 14** shows that using finer bundles reduces the standard deviation of the controlled water volume while its average does not significantly change. Therefore, sharper computational results having smaller deviations would be derived with a larger total number of bundles. However, this will not be always the case [79]. There would exist some optimal scaling relationship between the total numbers of sample paths and bundles, which is possibly problem-dependent and should be addressed with deeper stochastic analysis. In this paper, the stochastic grid bundling has been applied only to the space of inflow, which is an uncontrolled and is therefore not updated at each Picard iteration. We also tried bundling grids in the spaces involving the other variables $q$ and $V$ that are updated at each iteration; however, we encountered divergence or oscillation of numerical solutions during iterations. Updating the variables $q$ and $V$ implies updating the corresponding grid as well, suggesting the need for a more stable algorithm. We have not found such a criterion, which will be explored in future to better understand applicability of the stochastic grid bundling to solving coupled FBSDEs.

We finally discuss impacts of constraining the state variables, especially the constraint of the water volume using the singular variables (3) numerically implemented by the projection (45). All the computational cases generate almost unimodal probability densities of the controlled water volume having the mode far from the lower and upper boundaries of $\Omega_V$. This unimodal nature effectively prevents the sample paths from touching the boundaries of $\Omega_V$, which can be why the numerical computation here was completed in a stable manner. For example, at each time step at most 0.04%, 0.04%, 0.03%, and 0.06% of the sample paths touched the boundaries of $\Omega_V$ for the cases **(a)**, **(b)**, **(c)**, and **(d)** under the nominal parameter values, respectively. In this view, the projection is necessary for effective computation of the present stochastic control problem. Furthermore, we found that projecting the water volume using (45) was essential for almost all the our computation in this sub-section because of the finding that the Picard iteration, even with the relaxation, diverges and consequently meaningful numerical solution was not obtained at all. Bounding the state variables thus played at least the two roles: realization of physically reasonable states and stable numerical computation. Although the latter is a byproduct of the former, both would be important for considering more advanced mathematical modeling and control not only of dam and reservoir systems but also of other storage systems in future.

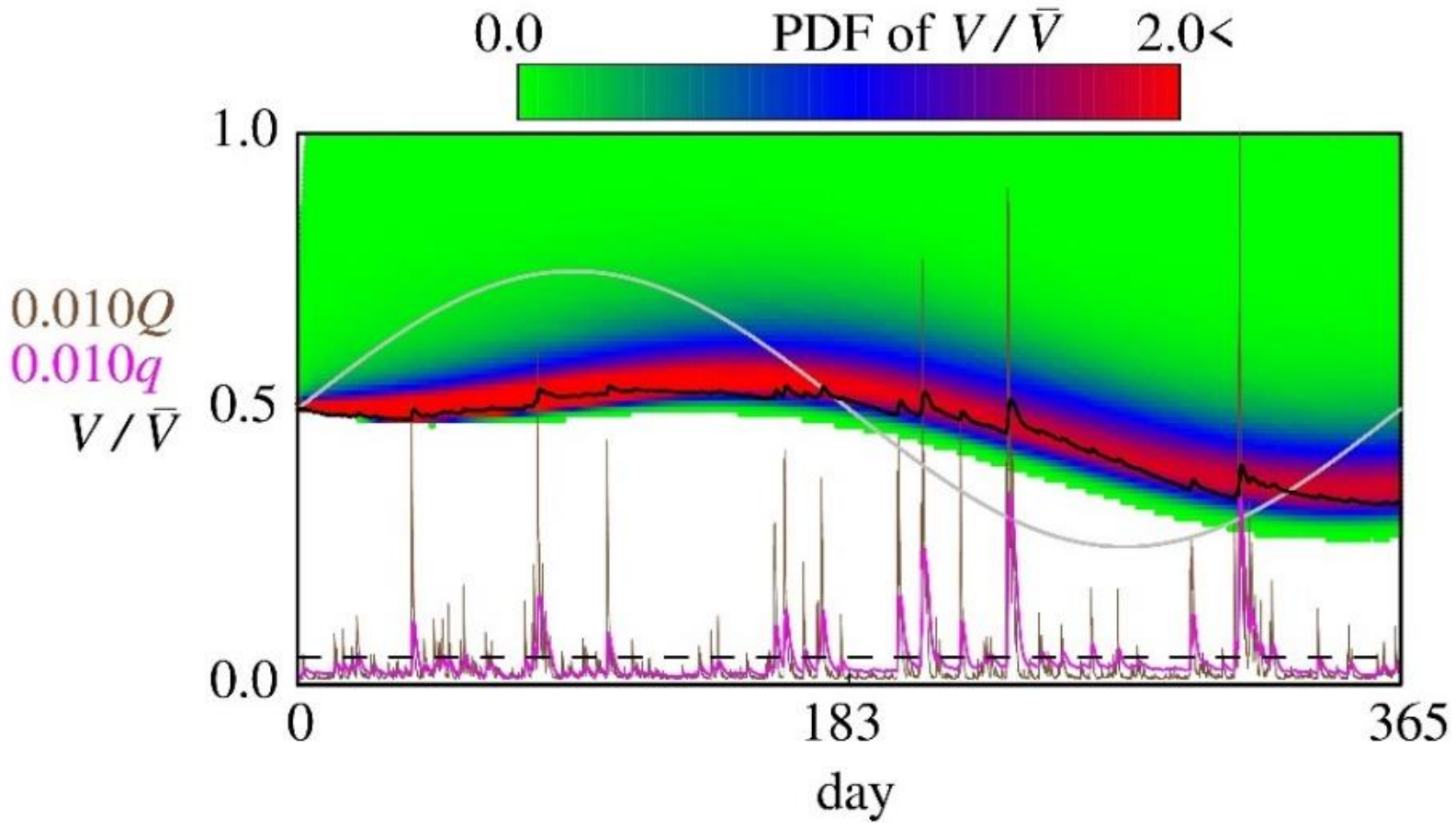

**Figure 4.** A sample path of the inflow (Brown line), the optimally controlled sample paths of the outflow (Purple line) and the water volume (Black line), and the corresponding probability density of the water volume (colored map) for the case **(a)**. The probability density is computed to be zero for white area. The broken line corresponds to the threshold $q = \bar{q}$ . The maximum $q$ was 104 ($m^3$/s).

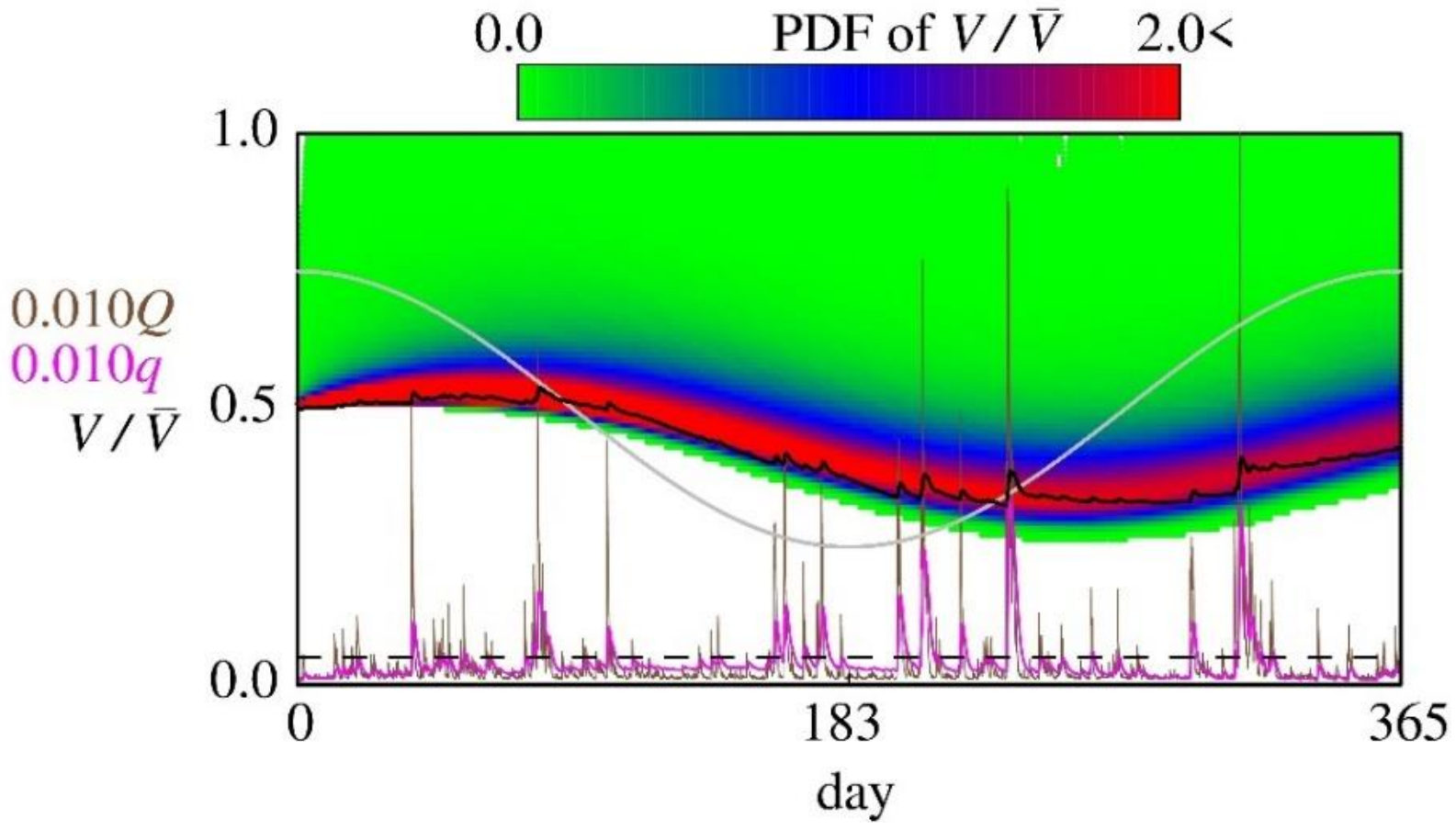

**Figure 5.** A set of sample paths and the corresponding probability density of the water volume for the case **(b)**. The same figure legends with **Figure 4**.

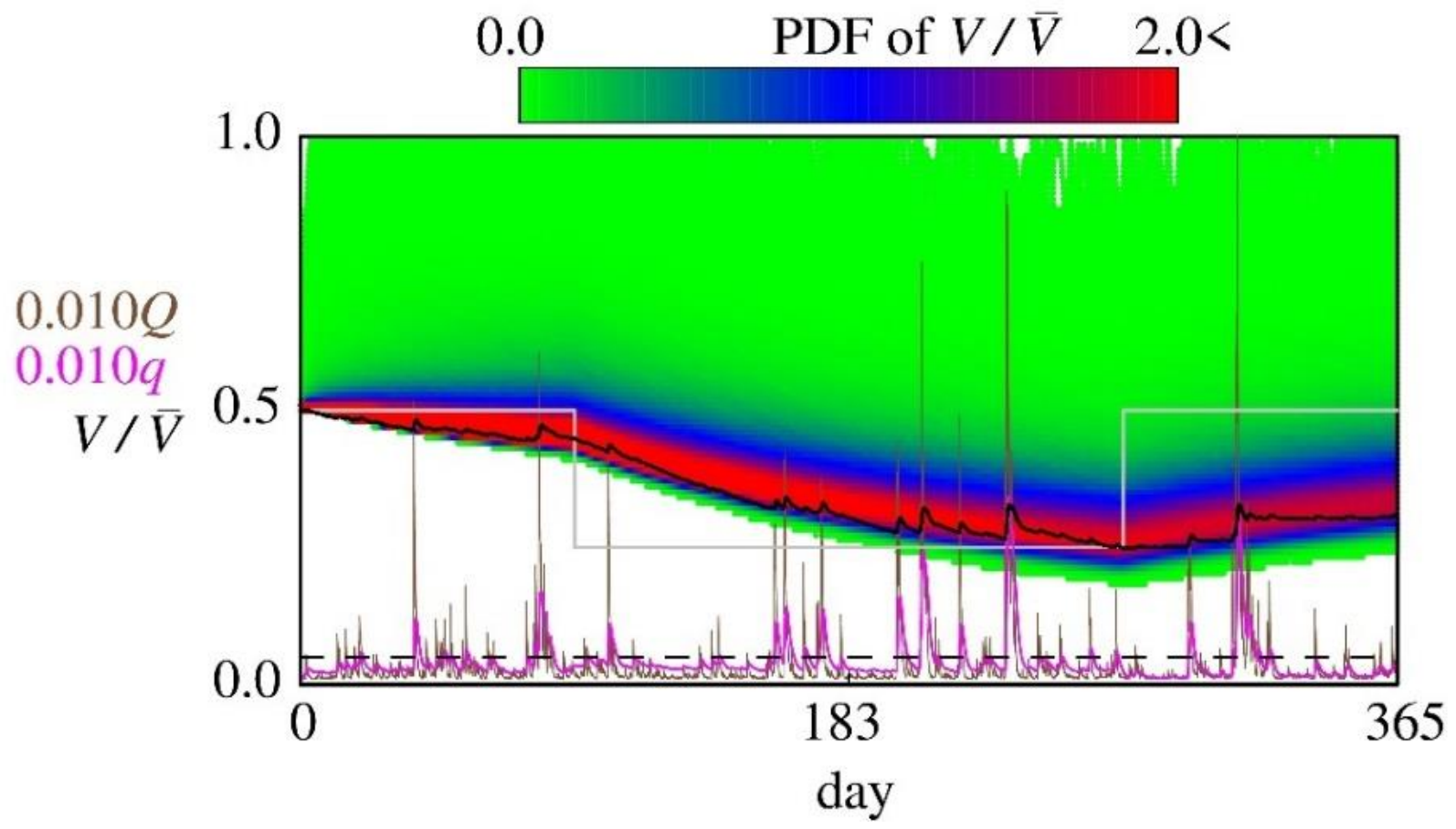

**Figure 6.** A set of sample paths and the corresponding probability density of the water volume for the case **(c)**. The same figure legends with **Figure 4**.

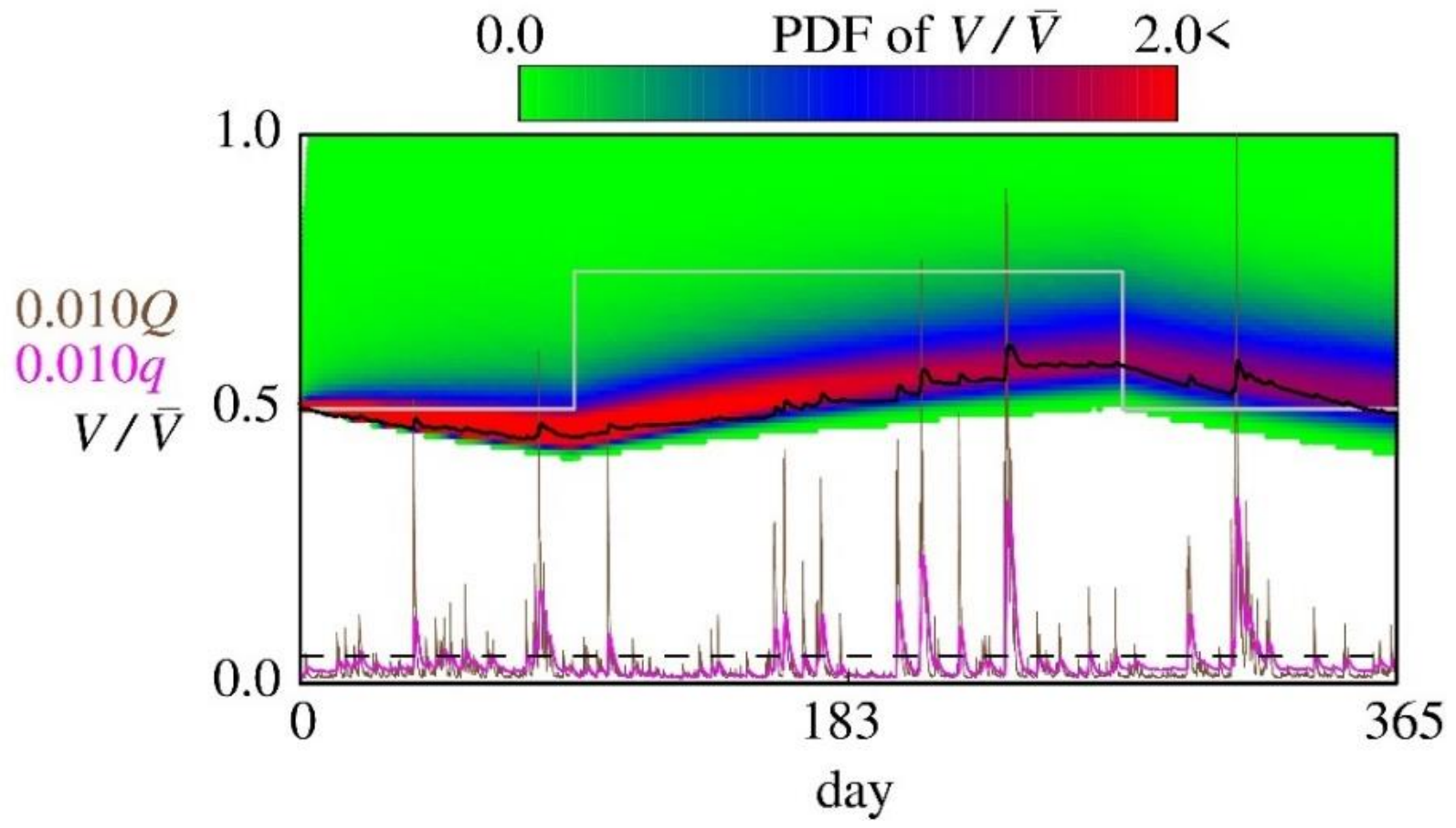

**Figure 7.** A set of sample paths and the corresponding probability density of the water volume for the case **(d)**. The same figure legends with **Figure 4**.

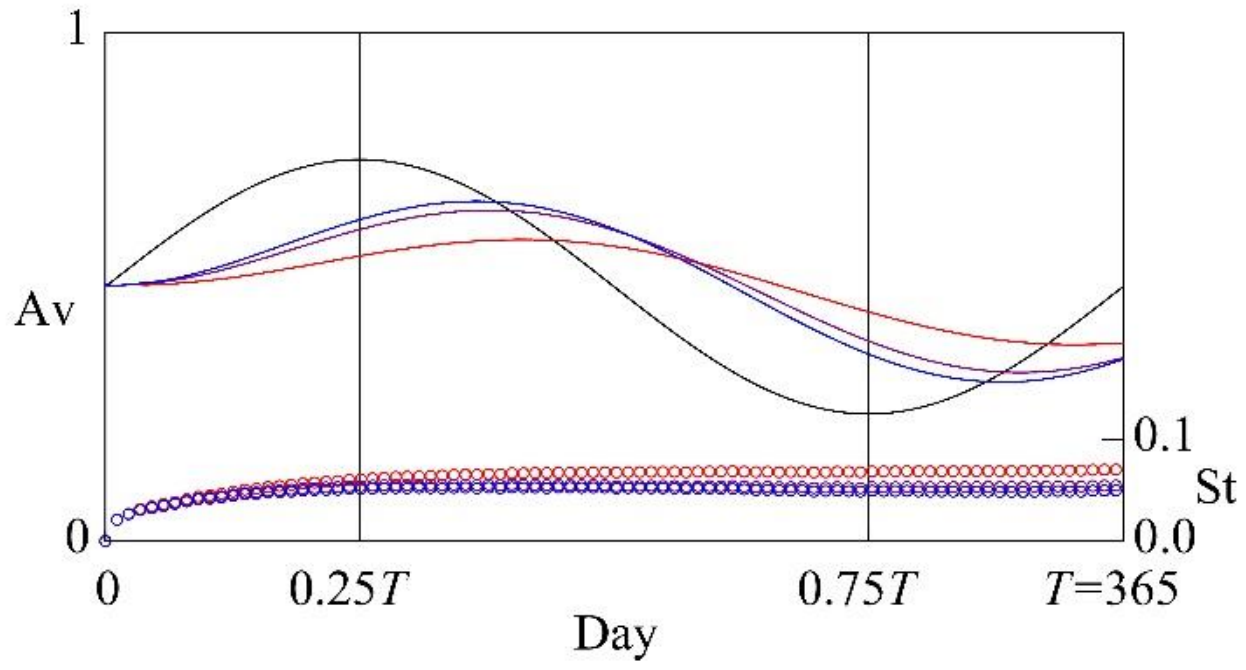

**Figure 8.** The target water volume (Black), and the averaged controlled water volumes (Av, plotted as curves) with the nominal $w_2$ (Red), 2.0 times of the nominal (Violet), and 2.5 times of the nominal (Blue). The standard deviations (St) of the controlled water volumes are also plotted with the same color legends (St, plotted as circles).

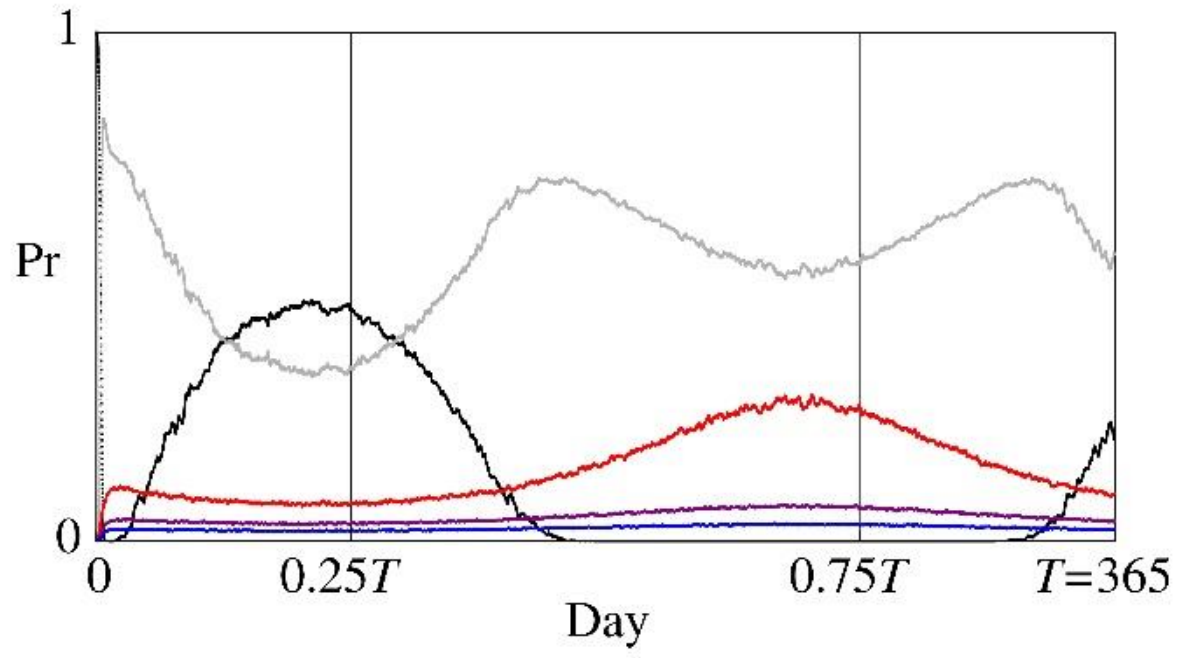

**Figure 9.** The computed probability (Pr) of the low outflow discharge $q$ for the case **(a)**. The color legends mean $\Pr(0 \le q < 2)$ (Black), $\Pr(2 \le q < 4)$ (Grey), $\Pr(4 \le q < 6)$ (Red), $\Pr(6 \le q < 8)$ (Violet), and $\Pr(8 \le q < 10)$ (Blue).

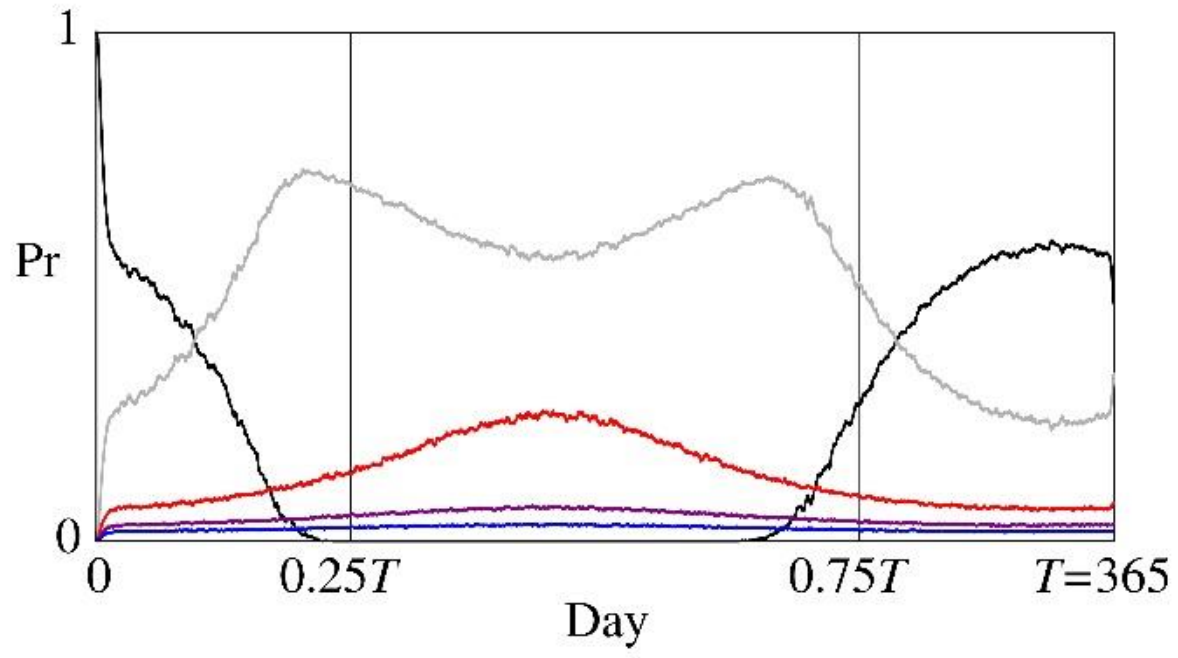

**Figure 10.** The computed probability (Pr) of the low outflow discharge $q$ for the case **(b)**. The same legends with **Figure 9**.

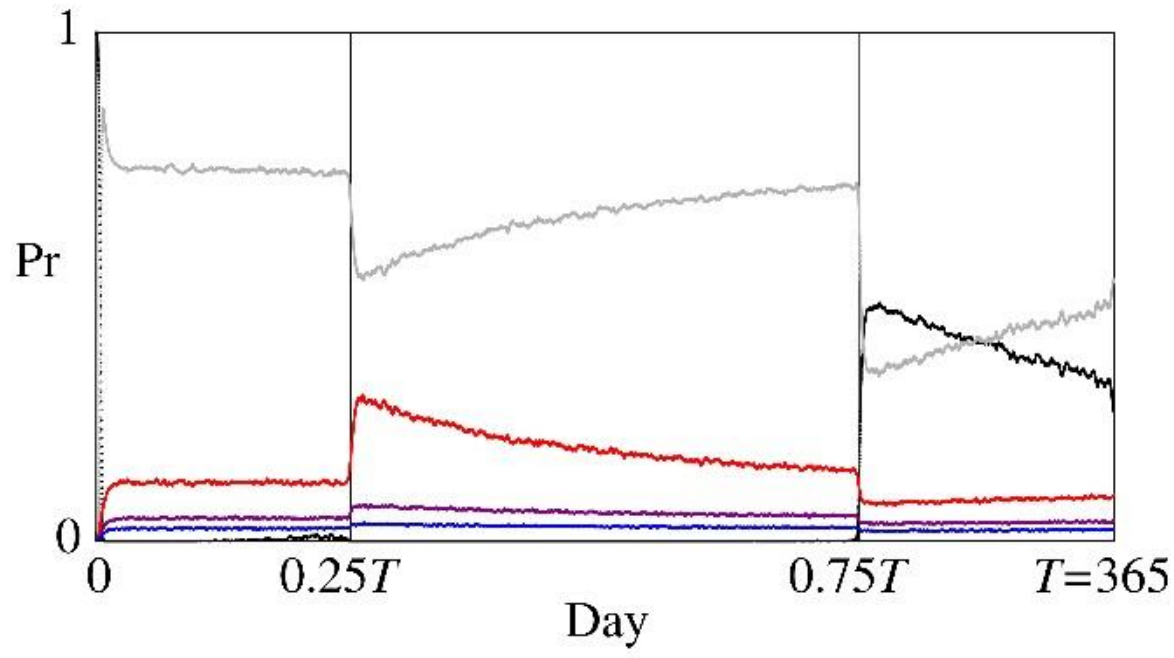

**Figure 11.** The computed probability (Pr) of the low outflow discharge $q$ for the case **(c)**. The same legends with **Figure 9.**

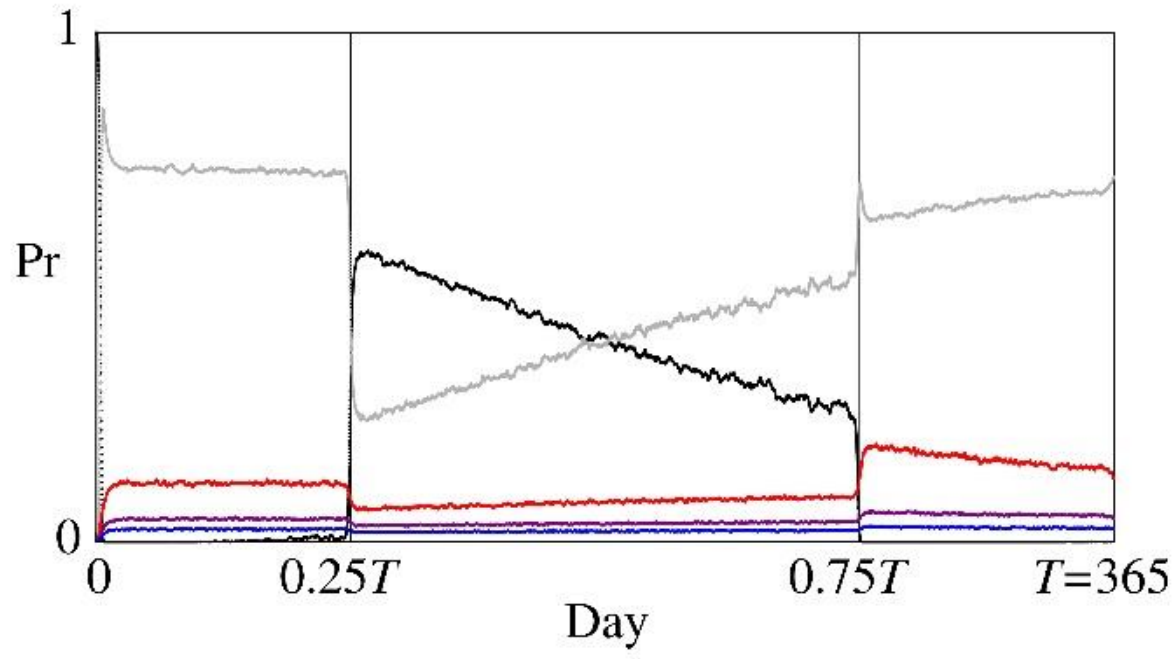

**Figure 12.** The computed probability (Pr) of the low outflow discharge $q$ for the case **(d)**. The same legends with **Figure 9**

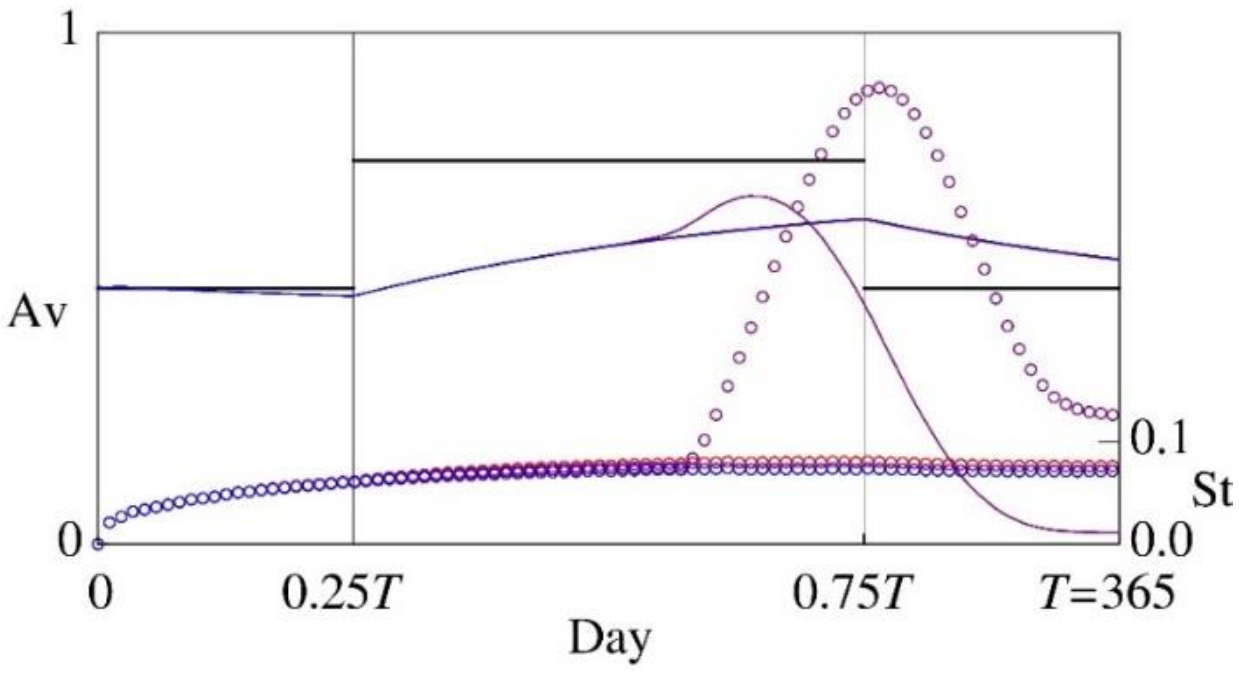

**Figure 13.** The target water volume (Black), and the averaged controlled water volumes (Av, plotted as curves) with the basis $\mathcal{S}_{\mathrm{LQ}}$ (Red), $\mathcal{S}_{\mathrm{NLQ1}}$ (Violet), $\mathcal{S}_{\mathrm{NLQ2}}$ (Blue). The standard deviations (St) of the controlled water volumes are also plotted with the same color legends (St, plotted as circles).

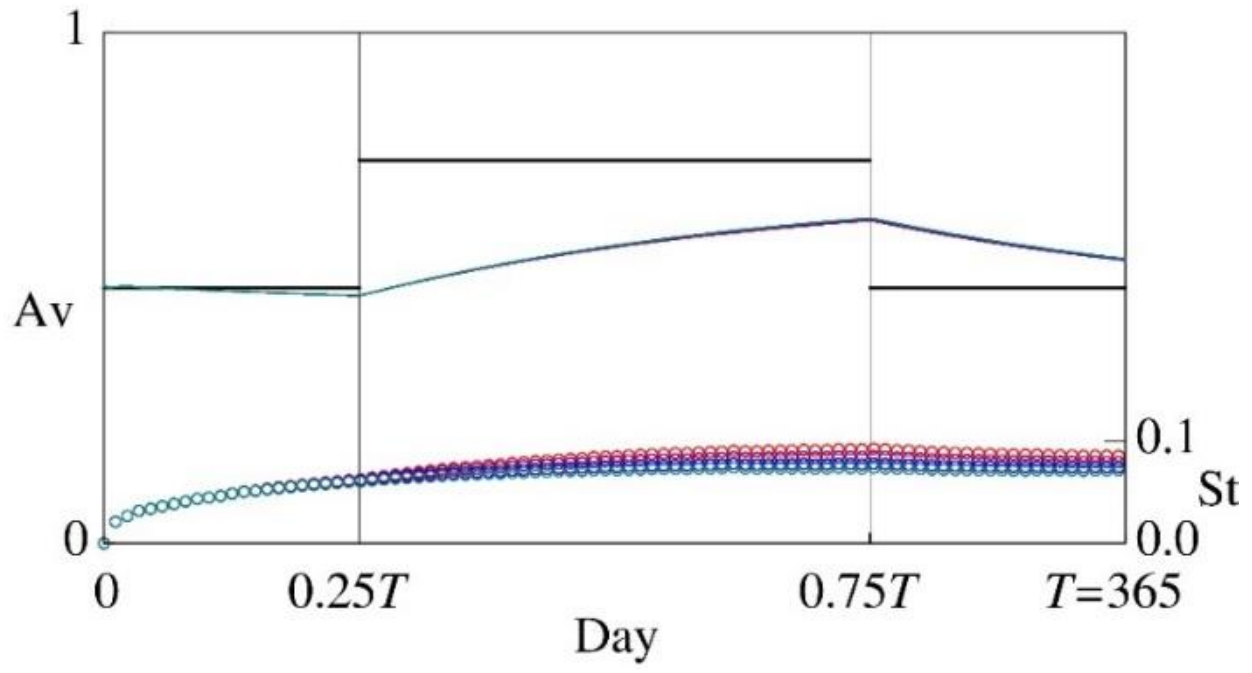

**Figure 14.** The target water volume (Black), and the averaged controlled water volumes (Av, plotted as curves) with the nominal $n_{\mathrm{bundle}} = 1$ (Red), $n_{\mathrm{bundle}} = 2$ (Violet), $n_{\mathrm{bundle}} = 4$ (Blue), and $n_{\mathrm{bundle}} = 8$ (Green). The standard deviations (St) of the controlled water volumes are also plotted with the same color legends (St, plotted as circles).

## 5. Conclusion and Perspectives

We proposed a new mathematical framework for controlling dam and reservoir systems based on a CBI process and FBSDEs. They were exactly solved in the LQ case. In addition, a sufficient optimality principle was obtained. We applied a least-squares Monte-Carlo method with a stochastic grid bundling and a relaxed Picard iteration to numerical computation of the FBSDEs. Computational examples based on an identified model suggested that the proposed approach can be used as a primitive for analysis of dam and reservoir systems using FBSDEs.

We obtained a series of mathematical and numerical results on the unique FBSDEs, but there are still issues to be tackled in future. In this paper, the constrained case was dealt with in a sub-optimal manner. One will improve this point by applying the Neumann boundary condition [66], which may arise in the stochastic maximum principle as constraints of adjoint variables. Optimizing the singular control variables [84-85] in a computationally feasible way is a non-trivial task. Using a soft-computing technique [86] or a penalization method [87] may mitigate the singularity, but convergence with respect to penalty and hyper parameters have to be carefully analyzed. A massive computational architecture will then be needed [41].

Another numerical issue to be considered in future is the choice of basis. This is a key issue when computing BSDEs and related stochastic optimization models [88]. We combined the stochastic grid bundling and the polynomial-like basis, but other choices such as Fourier expansions, Gaussian process regression, or deep learning technique [89] will be applicable. Computational efficiency and theoretical convergence should be considered simultaneously for choosing basis. For kernel ridge regression methods, Hu and Zastawniak [79] developed another stochastic grid bundling that works for high-dimensional problems, which can potentially cover control of dam and reservoir systems under wider conditions.

In general, solving coupled FBSDEs requires an iteration method as in this paper. However, its applicability is still limited by regularity of solutions to FBSDEs; especially, Picard iterations do not converge when the driver of the BSDE growths fast or the discount rate is small [40]. Employing a multi-level Picard iteration [90] may mitigate this issue.

From a theoretical viewpoint, there would be more than one decision-makers in one problem. Although we focused on a controlling problem of the outflow discharge, but real river management concerns balancing human activities, fisheries, and sediment storage in a dam-downstream river [91-92]. In such cases, decision-makers to control different state variables are not necessarily the same, leading us to consider a differential game. Another interesting case where there are more than one decision-makers arises if the coefficients are estimated only inaccurately [93] where the nature as the source of ambiguity is an opponent, leading us to formulate a differential game of the worst-case optimization type. The BSDE in this case part has been found to possess an exponential nonlinearity [24, 93], requiring more sophisticated methodologies both in mathematically and numerically. More realistic CBI processes for the streamflow dynamics would be found by considering a general form [94], but potentially requires a larger difficulty of model identification and faces with larger model ambiguity. We should therefore balance complexity and efficiency. We will tackle these issues along with continuous data collection in the study site.

## Appendix A: Motivating example

A motivating example of the proposed model is explained. We preliminarily found that the following operation rule, although it seems to be simple to represent real operation rules of dam and reservoir systems, work fairly well:

$$q_{(k+1)h} = \max\left\{0, q_{kh} + \left(C_1 + C_Q Q_{kh} + C_q q_{kh} + C_V V_{kh}\right)h\right\} \tag{56}$$

with non-negative integers $k$ and a time increment $h > 0$, and constants $C_1, C_Q, C_q, C_V$. In the Obara Dam case in **Section 4** the data with $h = 1$ (h) is available [77]. The operation rule (56) corresponds to the optimal control of the LQ case derived in **Proposition 2** because the acceleration of the outflow discharge linearly depends on the state variables. The non-negativity constraint in (56) is to avoid negative outflow discharge. Given time series data of $Q_{kh}$ and $V_{kh}$, we can simulate $q_{kh}$ using (56).

We fit the discrete dynamics (56) against the operation data of the OSO system where hourly time series of both $Q_{kh}$ and $V_{kh}$ are available. The data used here is the same with that in **Section 4.2** [77]. Using a least square regression of $q_{kh}$ (m$^3$/s) ($k = 1, 2, 3, \ldots$) between the model (56) and the data, we get the following parameter values: $C_1 = 0.3970$ (1/s/h), $C_Q = 0.2511$ (1/h), $C_q = -0.2812$ (1/h), and $C_V = -1.124 \times 10^{-9}$ (1/s/h). Using these parameter values, the $R^2$ coefficient of $q_{kh}$ between the data and the model from April 1 in 2016 to March 31 2020 is 0.86, suggesting that they agree well despite the simplicity of the model. Nevertheless, there is a discrepancy between them; the discrete model underestimates the real outflow (**Figure A1**). This underestimation implies that the model is too simple, especially lack of some mechanism to avoid too small outflow discharge like the last term of (11). The preliminary findings explained above motivated us to consider the model in the main text.

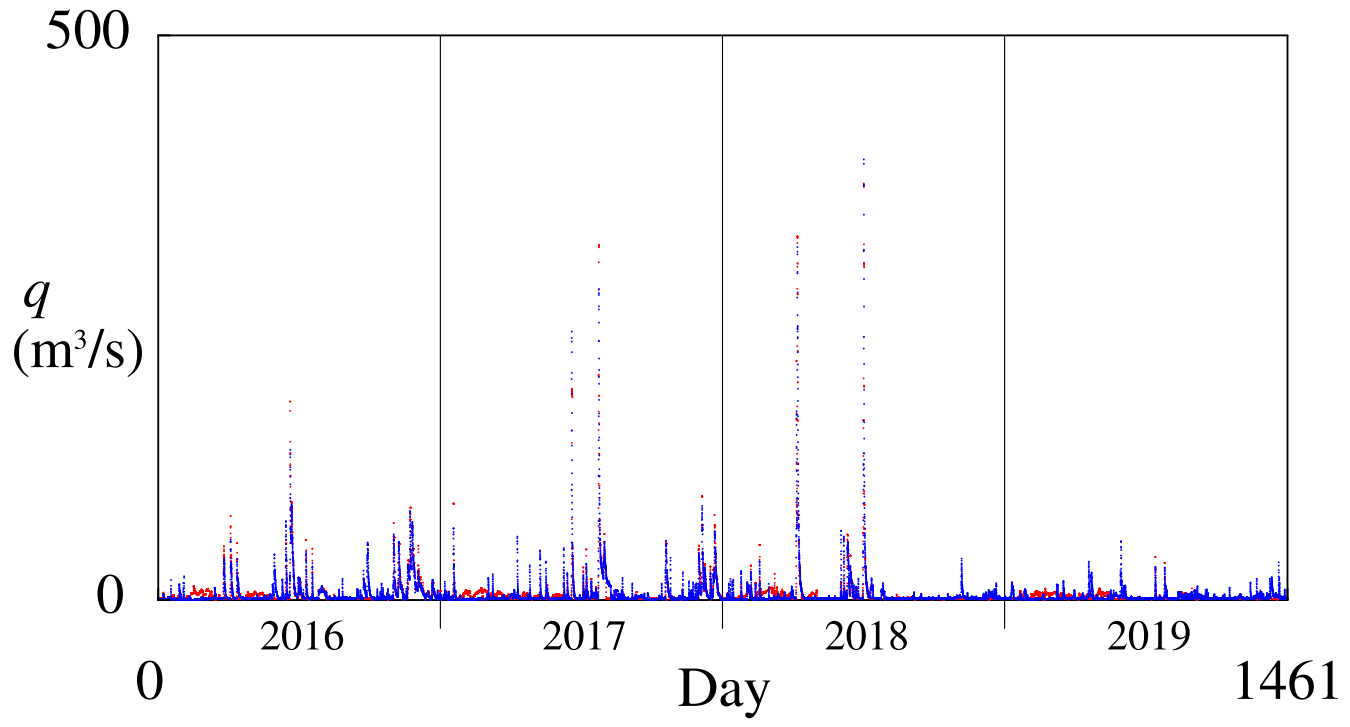

**Figure A1.** The observed outflow discharge (Red), and modelled outflow discharge (Blue). The modelled outflow underestimates the observed one at relatively low discharges.

## Appendix B

We present here analytical formulae of the statistical moments of the moments (Ave, Sta, Ske, Kur) of the inflow discharge $Q_t$ for a stationary state. This technique is the same with that for Hawkes models

(Lemma 4.4 of Bernis and Scotti [95]). For Ave, as in **Section 2**, we have

$$\text{Ave} = \left(1 - M_1\right)^{-1} \underline{Q} \ . \tag{57}$$

By applying Itô's formula to $\left(Q_t\right)^k$ ($k \in \mathbb{N}$) and the SDE (1) yields

$$\mathrm{d}\left(Q_t\right)^k = k\left(Q_t\right)^{k-1} \rho\left(\underline{Q} - Q_t\right)\mathrm{d}t + \left(Q_{t-} + z_t\right)^k - \left(Q_{t-}\right)^k , \tag{58}$$

where $z_t$ is the jump generated by $N$ at time $t$. For $k = 2$, from (58) we get

$$\mathrm{d}\left(Q_t\right)^2 = 2\rho\left\{\underline{Q}Q_t - \left(Q_t\right)^2\right\}\mathrm{d}t + \left(z_t\right)^2 + 2Q_{t-}z_t \ . \tag{59}$$

Taking the expectation of (59) yields

$$\mathrm{dE}\left[\left(Q_t\right)^2\right] = 2\rho\left\{\underline{Q}\mathrm{E}\left[Q_t\right] - \mathrm{E}\left[\left(Q_t\right)^2\right]\right\}\mathrm{d}t + \mathrm{E}\left[\left(z_t\right)^2\right] + 2\mathrm{E}\left[Q_{t-}z_t\right]. \tag{60}$$

Because the jump is generated by a state-dependent jump kernel $\dfrac{Q_{t-}\rho a}{z^{1+\alpha}} e^{-bz}\mathrm{d}z$, we formally have

$$\mathrm{E}\left[\left(z_t\right)^2\right] = \mathrm{E}\left[\int_0^\infty \frac{Q_{t-}\rho a}{z^{1+\alpha}} z^2 e^{-bz}\mathrm{d}z\right]\mathrm{d}t = \rho M_2 \mathrm{E}\left[Q_t\right]\mathrm{d}t \tag{61}$$

and

$$\mathrm{E}\left[Q_{t-}z_t\right] = \mathrm{E}\left[Q_{t-}\int_0^\infty \frac{Q_{t-}\rho a}{z^{1+\alpha}} z e^{-bz}\mathrm{d}z\right]\mathrm{d}t = \left(\int_0^\infty \frac{\rho a}{z^{\alpha}} e^{-bz}\mathrm{d}z\right)\mathrm{E}\left[\left(Q_{t-}\right)^2\right]\mathrm{d}t = \rho M_1 \mathrm{E}\left[\left(Q_t\right)^2\right]\mathrm{d}t , \tag{62}$$

where $M_k = \int_0^\infty \frac{a}{z^{\alpha+1}} z^k e^{-bz}\mathrm{d}z = b^{\alpha-k}\Gamma\left(k - \alpha\right)$ ($k \in \mathbb{N}$). By (60)-(62), we obtain

$$\frac{\mathrm{dE}\left[\left(Q_t\right)^2\right]}{\mathrm{d}t} = 2\rho\left\{\underline{Q}\mathrm{E}\left[Q_t\right] - \mathrm{E}\left[\left(Q_t\right)^2\right]\right\} + \rho M_2 \mathrm{E}\left[Q_t\right] + 2\rho M_1 \mathrm{E}\left[\left(Q_t\right)^2\right], \tag{63}$$

which reduces to the following algebraic equation under a stationary state:

$$2\underline{Q}\mathrm{E}\left[Q_t\right] - 2\mathrm{E}\left[\left(Q_t\right)^2\right] + M_2\mathrm{E}\left[Q_t\right] + 2M_1\mathrm{E}\left[\left(Q_t\right)^2\right] = 0 \tag{64}$$

Because we have $\text{Sta}^2 = \mathrm{E}\left[\left(Q_t\right)^2\right] - \text{Ave}^2 = \mathrm{E}\left[\left(Q_t\right)^2\right] - \left(1 - M_1\right)\underline{Q}^2$, from (64) we obtain

$$\text{Sta} = \sqrt{\frac{M_2}{2\left(1 - M_1\right)}\text{Ave}} \ . \tag{65}$$

Higher-order moments can also be obtained recursively from (58) although they are lengthy.

## Appendix C: Proofs of the propositions

***Proof of Proposition 1***

A solution to the FBSDEs containing (1)-(3), (20)-(22), and (26) is expressed as $\left(\tilde{Q},\tilde{q},\tilde{V},\tilde{p}^{(Q)},\tilde{p}^{(q)},\tilde{p}^{(V)},\tilde{\theta}^{(Q)},\tilde{\theta}^{(q)},\tilde{\theta}^{(V)}\right)$. Forward dynamics driven by $a \in \mathcal{A}_1$ is expressed as $\left(Q,q,V\right)$.

The proof here is based on that of Theorem 3.4 of Hess [31] but is different in several points. Firstly, the problem in the literature is to control a 1-D system, while ours is 3-D. Another difference is that a CBI process is controlled in the literature, while it is not controlled but serving as a driving noise process in our case; we therefore have $\tilde{Q}=Q$ . The latter difference is the reason the condition corresponding to the first boundedness assumption in Theorem 3.4 of Hess [31] is unnecessary.

Set the state and adjoint vectors

$$\mathbf{X}=\begin{pmatrix} Q \\ q \\ V \end{pmatrix}, \quad \tilde{\mathbf{X}}=\begin{pmatrix} \tilde{Q} \\ \tilde{q} \\ \tilde{V} \end{pmatrix}, \quad \tilde{\mathbf{p}}=\begin{pmatrix} \tilde{p}^{(Q)} \\ \tilde{p}^{(q)} \\ \tilde{p}^{(V)} \end{pmatrix}, \tag{66}$$

drift vectors,

$$\mathbf{b}=\begin{pmatrix} b^{(Q)} \\ b^{(q)} \\ b^{(V)} \end{pmatrix}=\begin{pmatrix} \rho\left\{\underline{Q}-(1-M_1)Q\right\} \\ a \\ Q-q \end{pmatrix}, \quad \mathbf{b}=\begin{pmatrix} \tilde{b}^{(Q)} \\ \tilde{b}^{(q)} \\ \tilde{b}^{(V)} \end{pmatrix}=\begin{pmatrix} \rho\left\{\underline{Q}-(1-M_1)\tilde{Q}\right\} \\ \tilde{a} \\ \tilde{Q}-\tilde{q} \end{pmatrix}, \tag{67}$$

and gradient vectors

$$\nabla J=\begin{pmatrix} \dfrac{\partial J}{\partial Q} \\ \dfrac{\partial J}{\partial q} \\ \dfrac{\partial J}{\partial V} \end{pmatrix} \quad \text{and} \quad \nabla H=\begin{pmatrix} \dfrac{\partial H}{\partial Q} \\ \dfrac{\partial H}{\partial q} \\ \dfrac{\partial H}{\partial V} \end{pmatrix}=\nabla J+\begin{pmatrix} \tilde{p}^{(Q)}\dfrac{\partial \tilde{b}^{(Q)}}{\partial Q}+\tilde{p}^{(q)}\dfrac{\partial \tilde{b}^{(q)}}{\partial Q}+\tilde{p}^{(V)}\dfrac{\partial \tilde{b}^{(V)}}{\partial Q} \\ \tilde{p}^{(Q)}\dfrac{\partial \tilde{b}^{(q)}}{\partial q}+\tilde{p}^{(q)}\dfrac{\partial \tilde{b}^{(q)}}{\partial q}+\tilde{p}^{(V)}\dfrac{\partial \tilde{b}^{(V)}}{\partial q} \\ \tilde{p}^{(Q)}\dfrac{\partial \tilde{b}^{(V)}}{\partial V}+\tilde{p}^{(q)}\dfrac{\partial \tilde{b}^{(q)}}{\partial V}+\tilde{p}^{(V)}\dfrac{\partial \tilde{b}^{(V)}}{\partial V} \end{pmatrix}+\begin{pmatrix} \int_0^\infty\left[\tilde{\theta}^{(Q)}(Q,z)+\tilde{\theta}^{(q)}(Q,z)+\tilde{\theta}^{(V)}(Q,z)\right]\nu(\mathrm{d}z) \\ 0 \\ 0 \end{pmatrix}, \tag{68}$$

where the subscript representing the time has been omitted for simplicity.

We prove the statement with the null terminal condition of the FBSDEs in mind. Firstly, we have

$$\phi\left(t,\tilde{Q}_t,\tilde{q}_t,\tilde{V}_t;\tilde{a}\right)-\phi\left(t,Q_t,q_t,V_t;a\right)=\mathbb{E}\left[\int_t^T e^{-\delta(s-t)}\left\{J\left(s,\tilde{Q}_s,\tilde{q}_s,\tilde{V}_s,\tilde{a}_s\right)-J\left(s,Q_s,q_s,V_s,a_s\right)\right\}\mathrm{d}s\middle|\mathcal{F}_t\right]. \tag{69}$$

By the definition of $H$ , we proceed as

$$
\begin{aligned}
&\mathbb{E}\left[\int_t^T e^{-\delta(s-t)}\left\{J\left(s,\tilde{Q}_s,\tilde{q}_s,\tilde{V}_s,\tilde{a}_s\right)-J\left(s,Q_s,q_s,V_s,a_s\right)\right\}\mathrm{d}s\middle|\mathcal{F}_t\right]\\
&=\mathbb{E}\left[\int_t^T e^{-\delta(s-t)}\left\{H\left(s,\tilde{Q}_s,\tilde{q}_s,\tilde{V}_s,\tilde{a}_s\right)-H\left(s,Q_s,q_s,V_s,a_s\right)\right\}\mathrm{d}s\middle|\mathcal{F}_t\right]\\
&-\mathbb{E}\left[\int_t^T e^{-\delta(s-t)}\left(\tilde{\mathbf{b}}_s-\mathbf{b}_s\right)\cdot\tilde{\mathbf{p}}_s\mathrm{d}s\middle|\mathcal{F}_t\right]\\
&+\mathbb{E}\left[\int_t^T e^{-\delta(s-t)}\delta\left(\tilde{\mathbf{X}}_s-\mathbf{X}_s\right)\cdot\tilde{\mathbf{p}}_s\mathrm{d}s\middle|\mathcal{F}_t\right]\\
&-\mathbb{E}\left[\int_t^T\int_0^\infty\int_{Q_{t-}}^{\tilde{Q}_{t-}} e^{-\delta(s-t)}\left\{\tilde{\theta}^{(Q)}(s,u,z)+\tilde{\theta}^{(q)}(s,u,z)+\tilde{\theta}^{(V)}(s,u,z)\right\}\mathrm{d}u\nu(\mathrm{d}z)\mathrm{d}s\middle|\mathcal{F}_t\right]\\
&=\mathbb{E}\left[\int_t^T e^{-\delta(s-t)}\left\{H\left(s,\tilde{Q}_s,\tilde{q}_s,\tilde{V}_s,\tilde{a}_s\right)-H\left(s,Q_s,q_s,V_s,a_s\right)\right\}\mathrm{d}s\middle|\mathcal{F}_t\right]\\
&-\mathbb{E}\left[\int_t^T e^{-\delta(s-t)}\left(\tilde{\mathbf{b}}_s-\mathbf{b}_s\right)\cdot\tilde{\mathbf{p}}_s\mathrm{d}s\middle|\mathcal{F}_t\right]+\mathbb{E}\left[\int_t^T e^{-\delta(s-t)}\delta\left(\tilde{\mathbf{X}}_s-\mathbf{X}_s\right)\cdot\tilde{\mathbf{p}}_s\mathrm{d}s\middle|\mathcal{F}_t\right]
\end{aligned}
\quad , \tag{70}
$$

where we used $\tilde{Q}\equiv Q$ and omitted the arguments $\left(\tilde{p}^{(Q)},\tilde{p}^{(q)},\tilde{p}^{(V)},\tilde{\theta}^{(Q)},\tilde{\theta}^{(q)},\tilde{\theta}^{(V)}\right)$ from $H$ for simplicity. By a co-variation argument (Appendix of Hess [31] for the noise part and Maslowski and Veverka [96] for the discount part), we obtain

$$
\begin{aligned}
\mathbb{E}\left[\left(\tilde{\mathbf{X}}_T-\mathbf{X}_T\right)\cdot\tilde{\mathbf{p}}_T\middle|\mathcal{F}_t\right]&=\mathbb{E}\left[\int_t^T e^{-\delta(s-t)}\left(\tilde{\mathbf{b}}_s-\mathbf{b}_s\right)\cdot\tilde{\mathbf{p}}_s\mathrm{d}s\middle|\mathcal{F}_t\right]\\
&-\mathbb{E}\left[\int_t^T e^{-\delta(s-t)}\delta\left(\tilde{\mathbf{X}}_s-\mathbf{X}_s\right)\cdot\tilde{\mathbf{p}}_s\mathrm{d}s\middle|\mathcal{F}_t\right]\\
&-\mathbb{E}\left[\int_t^T e^{-\delta(s-t)}\left(\tilde{\mathbf{X}}_s-\mathbf{X}_s\right)\cdot\nabla H\mathrm{d}s\middle|\mathcal{F}_t\right]
\end{aligned}
\tag{71}
$$

as well as $\mathbb{E}\left[\left(\tilde{\mathbf{X}}_T-\mathbf{X}_T\right)\cdot\tilde{\mathbf{p}}_T\middle|\mathcal{F}_t\right]=0$ by the terminal condition. Combining (69)-(71) yields the identity

$$
\begin{aligned}
&\phi\left(t,\tilde{Q}_t,\tilde{q}_t,\tilde{V}_t;\tilde{a}\right)-\phi\left(t,Q_t,q_t,V_t;a\right)\\
&=\mathbb{E}\left[\int_t^T e^{-\delta(s-t)}\left\{H\left(s,\tilde{Q}_s,\tilde{q}_s,\tilde{V}_s,\tilde{a}_s\right)-H\left(s,Q_s,q_s,V_s,a_s\right)\right\}\mathrm{d}s\middle|\mathcal{F}_t\right]\\
&-\mathbb{E}\left[\int_t^T e^{-\delta(s-t)}\left(\tilde{\mathbf{X}}_s-\mathbf{X}_s\right)\cdot\nabla H\mathrm{d}s\middle|\mathcal{F}_t\right]
\end{aligned}
. \tag{72}
$$

By the convexity of $H$ and the maximizing property of $\tilde{a}$, we obtain

$$
\begin{aligned}
&\mathbb{E}\left[\int_t^T e^{-\delta(s-t)}\left\{H\left(s,\tilde{Q}_s,\tilde{q}_s,\tilde{V}_s,\tilde{a}_s\right)-H\left(s,Q_s,q_s,V_s,a_s\right)\right\}\mathrm{d}s\middle|\mathcal{F}_t\right]\\
&=\mathbb{E}\left[\int_t^T e^{-\delta(s-t)}\left\{H\left(s,\tilde{Q}_s,\tilde{q}_s,\tilde{V}_s,\tilde{a}_s\right)-H\left(s,Q_s,q_s,V_s,\tilde{a}_s\right)\right\}\mathrm{d}s\middle|\mathcal{F}_t\right]\\
&+\mathbb{E}\left[\int_t^T e^{-\delta(s-t)}\left\{H\left(s,Q_s,q_s,V_s,\tilde{a}_s\right)-H\left(s,Q_s,q_s,V_s,a_s\right)\right\}\mathrm{d}s\middle|\mathcal{F}_t\right]\\
&\geq\mathbb{E}\left[\int_t^T e^{-\delta(s-t)}\left\{H\left(s,\tilde{Q}_s,\tilde{q}_s,\tilde{V}_s,\tilde{a}_s\right)-H\left(s,Q_s,q_s,V_s,\tilde{a}_s\right)\right\}\mathrm{d}s\middle|\mathcal{F}_t\right]\\
&\geq\mathbb{E}\left[\int_t^T e^{-\delta(s-t)}\left(\tilde{\mathbf{X}}_s-\mathbf{X}_s\right)\cdot\nabla H\mathrm{d}s\middle|\mathcal{F}_t\right]
\end{aligned}
. \tag{73}
$$

By (72)-(73), we get the desired optimality because $a\in\mathcal{A}_1$ is arbitrary:

$$
\Phi\left(t,\tilde{Q}_t,\tilde{q}_t,\tilde{V}_t\right)=\phi\left(t,\tilde{Q}_t,\tilde{q}_t,\tilde{V}_t;\tilde{a}\right)\geq\phi\left(t,Q_t,q_t,V_t;a\right), \tag{74}
$$

□

***Proof of Proposition 2***

The time-dependent coefficients $A, B, C, D, E, F, G, I, J, K, L, O$ are unknowns having the terminal value 0. Below, we use the notation $A' = \frac{\mathrm{d}A}{\mathrm{d}t}$ etc. The argument $t$ is omitted here for simplicity.

On the one hand, based on the ansatz (29)-(31), we have

$$\begin{aligned}
\mathrm{d}p^{(Q)} &= \mathrm{d}\left(AQ + Bq + CV + D\right) \\
&= \left(A'Q + B'q + C'V + D'\right)\mathrm{d}t + \left(A\mathrm{d}Q + B\mathrm{d}q + C\mathrm{d}V\right) \\
&= \left(A' - \rho\left(1 - M_1\right)A + \frac{1}{w_3}BE + C\right)Q\mathrm{d}t + \left(B' - C + \frac{1}{w_3}BF\right)q\mathrm{d}t \\
&\quad + \left(C' + \frac{1}{w_3}BG\right)V\mathrm{d}t + \left(D' + \rho A\underline{Q} + \frac{1}{w_3}BI\right)\mathrm{d}t \\
&\quad + A\int_0^\infty \int_0^{Q_{t-}} z\tilde{N}\left(\mathrm{d}u, \mathrm{d}z, \mathrm{d}t\right)
\end{aligned} \quad . \tag{75}$$

On the other hand, we also have

$$\begin{aligned}
\mathrm{d}p^{(Q)} &= -\left\{-w_1\left(Q - q\right) - \left(\rho\left(1 - M_1\right) + \delta\right)p^{(Q)} + p^{(V)}\right\}\mathrm{d}t + \int_0^\infty \int_0^{Q_{t-}} \theta^{(Q)}\left(u, z\right)\tilde{N}\left(\mathrm{d}u, \mathrm{d}z, \mathrm{d}t\right) \\
&\quad + \int_0^\infty \left[\theta^{(Q)}\left(Q, z\right) + \theta^{(q)}\left(Q, z\right) + \theta^{(V)}\left(Q, z\right)\right]v\left(\mathrm{d}z\right)\mathrm{d}t \\
&= \left(w_1 + \left(\rho\left(1 - M_1\right) + \delta\right)A - J\right)Q\mathrm{d}t \\
&\quad + \left(-w_1 + \left(\rho\left(1 - M_1\right) + \delta\right)B - K\right)q\mathrm{d}t + \left(\left(\rho\left(1 - M_1\right) + \delta\right)C - L\right)V\mathrm{d}t \\
&\quad + \left(\left(\rho\left(1 - M_1\right) + \delta\right)D - O\right)\mathrm{d}t \\
&\quad + \int_0^\infty \left[\theta^{(Q)}\left(Q, z\right) + \theta^{(q)}\left(Q, z\right) + \theta^{(V)}\left(Q, z\right)\right]v\left(\mathrm{d}z\right)\mathrm{d}t + \int_0^\infty \int_0^{Q_{t-}} \theta^{(Q)}\left(u, z\right)\tilde{N}\left(\mathrm{d}u, \mathrm{d}z, \mathrm{d}t\right)
\end{aligned} \quad . \tag{76}$$

Equating the coefficients between (75)-(76) yields the ODEs for $A, B, C, D$ as follows. We have

$$\begin{aligned}
&\left(A' - \rho\left(1 - M_1\right)A + \frac{1}{w_3}BE + C\right)Q\mathrm{d}t + \left(B' - C + \frac{1}{w_3}BF\right)q\mathrm{d}t \\
&+ \left(C' + \frac{1}{w_3}BG\right)V\mathrm{d}t + \left(D' + \rho\underline{Q}A + \frac{1}{w_3}BI\right)\mathrm{d}t \\
&+ \int_0^\infty \int_0^{Q_{t-}} Az\tilde{N}\left(\mathrm{d}u, \mathrm{d}z, \mathrm{d}t\right) + \int_0^\infty \left[\theta^{(Q)}\left(Q, z\right) + \theta^{(q)}\left(Q, z\right) + \theta^{(V)}\left(Q, z\right)\right]v\left(\mathrm{d}z\right)\mathrm{d}t \\
&= \left(w_1 + \left(\rho\left(1 - M_1\right) + \delta\right)A - J\right)Q\mathrm{d}t \\
&+ \left(-w_1 + \left(\rho\left(1 - M_1\right) + \delta\right)B - K\right)q\mathrm{d}t + \left(\left(\rho\left(1 - M_1\right) + \delta\right)C - L\right)V\mathrm{d}t \\
&+ \left(\left(\rho\left(1 - M_1\right) + \delta\right)D - O\right)\mathrm{d}t + \int_0^\infty \int_0^{Q_{t-}} \theta^{(Q)}\left(u, z\right)\tilde{N}\left(\mathrm{d}u, \mathrm{d}z, \mathrm{d}t\right)
\end{aligned} \tag{77}$$

and thus

$$A' - \rho\left(1 - M_1\right)A + \frac{1}{w_3}BE + C = w_1 + \left(\rho\left(1 - M_1\right) + \delta\right)A - J \;, \tag{78}$$

$$B' - C + \frac{1}{w_3} BF = -w_1 + \left(\rho\left(1 - M_1\right) + \delta\right) B - K \ , \tag{79}$$

$$C' + \frac{1}{w_3} BG = \left(\rho\left(1 - M_1\right) + \delta\right) C - L \ , \tag{80}$$

$$D' + \rho \underline{Q} A + \frac{1}{w_3} BI + \int_0^\infty \left(\phi^{(Q)} + \phi^{(q)} + \phi^{(V)}\right) z v\left(\mathrm{d}z\right) = \left(\rho\left(1 - M_1\right) + \delta\right) D - O \ , \tag{81}$$

where we assumed

$$\left(\theta^{(Q)}\left(u, z\right), \theta^{(q)}\left(u, z\right), \theta^{(V)}\left(u, z\right)\right) = \left(\phi^{(Q)} z, \phi^{(q)} z, \phi^{(V)} z\right) . \tag{82}$$

We also obtain

$$\theta^{(Q)}\left(u, z\right) = Az \quad \text{and thus} \quad \phi^{(Q)} = A . \tag{83}$$

In an essentially the same way, for the BSDE of $p^{(q)}$, we get

$$\begin{aligned} & \left( E' - \rho\left(1 - M_1\right) E + \frac{1}{w_3} EF + G \right) Q\mathrm{d}t + \left( F' + \frac{1}{w_3} F^2 - G \right) q\mathrm{d}t + \left( G' + \frac{1}{r} FG \right) V\mathrm{d}t \\ & + \left( I' + \rho E \underline{Q} + \frac{1}{r} FI \right) \mathrm{d}t + \int_0^\infty Ez\tilde{N}\left(\mathrm{d}u, \mathrm{d}z, \mathrm{d}t\right) \\ & = \left(-w_1 + \delta E + J\right) Q\mathrm{d}t + \left(w_1 + \delta F + K\right) q\mathrm{d}t + \left(\delta G + L\right) V\mathrm{d}t + \left(\delta I + O\right) \mathrm{d}t \\ & + \int_0^\infty \int_0^{Q_{t-}} \theta^{(q)}\left(u, z\right) \tilde{N}\left(\mathrm{d}u, \mathrm{d}z, \mathrm{d}t\right) \end{aligned} \tag{84}$$

and

$$\theta^{(q)}\left(u, z\right) = Ez \quad \text{and thus} \quad \phi^{(q)} = E \ . \tag{85}$$

Then, we obtain the ODEs

$$E' - \rho\left(1 - M_1\right) E + \frac{1}{w_3} EF + G = -w_1 + \delta E + J \ , \tag{86}$$

$$F' + \frac{1}{w_3} F^2 - G = w_1 + \delta F + K \ , \tag{87}$$

$$G' + \frac{1}{w_3} FG = \delta G + L \ , \tag{88}$$

$$I' + \rho E \underline{Q} + \frac{1}{w_3} FI = \delta I + O \ . \tag{89}$$

Similarly, for $p^{(V)}$, we get

$$\begin{aligned} & \left( J' - \rho\left(1 - M_1\right) J + \frac{1}{w_3} EK + L \right) Q\mathrm{d}t + \left( K' + \frac{1}{w_3} KF - L \right) q\mathrm{d}t + \left( L' + \frac{1}{w_3} GK \right) V\mathrm{d}t \\ & + \left( O' + \rho \underline{Q} J + \frac{1}{w_3} IK \right) \mathrm{d}t + \int_0^\infty \int_0^{Q_{t-}} Jz\tilde{N}\left(\mathrm{d}u, \mathrm{d}z, \mathrm{d}t\right) \\ & = \delta JQ\mathrm{d}t + \delta Kq\mathrm{d}t + \left(w_2 + \delta L\right) V\mathrm{d}t + \left(-w_2 \hat{V} + \delta O\right) \mathrm{d}t \\ & + \int_0^\infty \int_0^{Q_{t-}} \theta^{(V)}\left(u, z\right) \tilde{N}\left(\mathrm{d}u, \mathrm{d}z, \mathrm{d}t\right) \end{aligned} \tag{90}$$

and

$$\theta^{(V)}(u,z) = Jz \quad \text{and thus} \quad \phi^{(V)} = J \,. \tag{91}$$

Then, we get the ODEs

$$J' - \rho(1-M_1)J + \frac{1}{w_3}EK + L = \delta J \,, \tag{92}$$

$$K' + \frac{1}{w_3}KF - L = \delta K \,, \tag{93}$$

$$L' + \frac{1}{w_3}GK = w_2 + \delta L \,, \tag{94}$$

$$O' + \rho \underline{Q} J + \frac{1}{w_3}IK = -w_2\hat{V} + \delta O \,. \tag{95}$$

Note that we have the equations

$$E' = -w_1 + \left(\rho(1-M_1)+\delta\right)E + J - G - \frac{1}{w_3}EF \quad \text{and} \quad B' = -w_1 + \left(\rho(1-M_1)+\delta\right)B + C - K - \frac{1}{w_3}BF \,, \tag{96}$$

$$J' = \left(\rho(1-M_1)+\delta\right)J - L - \frac{1}{w_3}EK \quad \text{and} \quad C' = \left(\rho(1-M_1)+\delta\right)C - L - \frac{1}{w_3}BG \,, \tag{97}$$

$$K' = \delta K + L - \frac{1}{w_3}KF \quad \text{and} \quad G' = \delta G + L - \frac{1}{w_3}FG \,. \tag{98}$$

They invoke the symmetry $E = B$, $J = C$, $K = G$. In summary, the system of ODEs to be solved contains nine equations subject to the homogenous terminal condition 0 at $t = T$, which are (33)-(41). The proof is then completed.

□

**Acknowledgements**
Kurita Water and Environment Foundation 19B018 and 20K004 and Grant for Environmental Research Projects from the Sumitomo Foundation 203160 support this research.

**Contributions**
HY: Authorization, Conceptualization, Methodology, Formal analysis, Computation, Review, Editing